\documentclass[final,3p,times,twocolumn]{elsarticle}
\usepackage{amssymb}
\usepackage{amsmath}

\usepackage{graphicx, subfigure}
\usepackage{soul}

\usepackage[%
backref=true,%
pagebackref=true,%
hyperfootnotes=true,%
hyperindex=true,%
pageanchor=true,%
plainpages=false,%
colorlinks=true,%  
linkcolor=blue,%
citecolor=red,%
urlcolor=cyan,%
linktocpage=true,% 
pdfstartview={Fit},
bookmarksopen=false
]{hyperref}
\usepackage{color}
\usepackage{hyperref}
\let\oldhref\href
\renewcommand{\href}[2]{\oldhref{#1}{\hbox{#2}}}
\journal{Astroparticle Physics}

\begin{document}

\begin{frontmatter}

\title{On the potential of Cherenkov Telescope Arrays and KM3 Neutrino Telescopes for the detection of extended sources}
 \author[gssi,mpik]{L. Ambrogi}  \ead{lucia.ambrogi@gssi.infn.it}
  \author[gssi]{S. Celli} \ead{silvia.celli@gssi.infn.it}
   \author[gssi,dias,mpik]{F. Aharonian}
 \address[gssi]{Gran Sasso Science Institute, viale Francesco Crispi, 7 67100 L'Aquila (AQ), Italy} 
 \address[dias]{Dublin Institute of Advanced Studies, 10 Burlington Road, Dublin 4, Ireland} 
 \address[mpik]{Max-Planck-Institut f\"ur Kernphysik, Saupfercheckweg 1, D-69117 Heidelberg, Germany} 

\begin{abstract}
We discuss the discovery potential of extended very-high-energy (VHE) neutrino sources by the future KM3 Neutrino Telescope (KM3NeT) in the context of the constraining power of the Cherenkov Telescope Array (CTA), designed for deep surveys of the sky in VHE gamma rays. The study is based on a comparative analysis of sensitivities of KM3NeT and CTA. We show that a minimum gamma-ray energy flux of $E^2 \phi_\gamma(10~\textrm{TeV})>1 \times 10^{-12}$~TeV~cm$^{-2}$~s$^{-1}$ is required to identify a possible neutrino counterpart with a 3$\sigma$ significance and 10 years of KM3NeT observations with upgoing muons, if the source has an angular size of $R_{src}=0.1$~deg and emits gamma rays with an $E^{-2}$ energy spectrum through a full hadronic mechanism. This minimum gamma-ray flux is increased to the level of $E^2 \phi_\gamma(10~\textrm{TeV})>2 \times 10^{-11}$~TeV~cm$^{-2}$~s$^{-1}$ in case of sources with radial extension of $R_{src}=2.0$~deg.
The analysis methods are applied to the supernova remnant RX J1713.7-3946 and the Galactic Center Ridge, as well as to the recent HAWC catalog of multi-TeV gamma-ray sources.
\end{abstract}


\end{frontmatter}

%%%%%%%%%%%%%%%%%%%%%%%%%%%%%%%%%%%%%%%%%%%''''
\section{Introduction}
\label{sec:Intro}
%%%%%%%%%%%%%%%%%%%%%%%%%%%%%%%%%%%%%%%%%%%''''
The progress made in ground-based gamma-ray astronomy over the last two decades has lead to the detection of more than 200 very-high-energy ($E \geq 200$~GeV) sources reported by the H.E.S.S. \cite{2015arXiv150903544C}, MAGIC \cite{Aleksic:2014lkm}, VERITAS \cite{Holder:2016heb} and HAWC \cite{Abeysekara:2013tza} collaborations. 
Recently, success has been reported also in neutrino astronomy by the detection of  a diffuse flux of multi-TeV neutrinos of extraterrestial origin by the IceCube collaboration \cite{ICdiscovery}. In the feasible future, the upgraded IceCube and the planned KM3NeT neutrino telescopes will serve as the major tools of VHE neutrinos. Apparently, the identification of objects contributing to the reported diffuse neutrino flux, as well as the discovery of discrete sources of VHE neutrinos is the major objective of neutrino astronomy for the coming years. So far, no clear association between any class of astrophysical sources and cosmic neutrinos has been identified. 
Despite the broad class of potential neutrino sources and the different possible scenarios of neutrino production in astrophysical environments, the production mechanisms of VHE neutrinos are connected to the hadronic interactions of ultra-relativistic protons with the ambient gas and radiation.  Namely,  the major production channels of VHE neutrinos are the decays of  charged $\pi^\pm$-mesons, the secondary products of hadronic $pp$ and $p \gamma$ interactions.  Since these processes are accompanied by the production and decay of $\pi^0$-mesons, the VHE gamma-rays and neutrinos are produced at comparable rates. Consequently, one would expect similar fluxes of gamma-rays and neutrinos. On the other hand,  the ground-based gamma-ray detectors, in particular the current arrays of Imaging Atmospheric Cherenkov Telescopes (IACT), H.E.S.S., MAGIC and VERITAS, provide lower flux sensitivities for point-like sources around $1$~TeV, compared to the sensitivities of the present IceCube and the forthcoming KM3NeT neutrino detectors. This circumstance reduces the chances of detection of discrete VHE neutrino sources, except for compact objects or sources located at cosmological distances. TeV gamma-ray fluxes from these objects are indeed expected to be suppressed because of both internal and intergalactic absorption, through photon-photon pair production interactions. In this regard, hidden sources constitute an interesting possibility for the explanation of the measured IceCube neutrino flux: these cosmic-ray accelerators, being surrounded by very dense environments, cannot be probed by gamma rays, while transparent to neutrinos. Among them, choked GRBs and supermassive black hole cores have widely been discussed in literature \cite{1981berez,2016murase, 2016senno}: in these cases, neutrinos constitute crucial probes in shedding light on the central engine activity. Otherwise, the VHE gamma-ray fluxes should be taken as a robust criterion regarding the expectations of discovery of discrete VHE neutrino sources. 
Given the difference in the TeV flux sensitivities of IACT arrays and KM3-scale neutrino detectors, the gamma-ray fluxes are especially constraining for point-like sources.  In this paper, a point-like source is called an astronomical object with angular extension less than the typical angular resolution of IACT arrays ($\sim 0.1^\circ$). 
For mildly-extended sources with an angular size $\sim 1^\circ$, which is one order of magnitude larger than the point spread function (PSF) of IACTs but still comparable to the PSF of VHE neutrino detectors, the  gamma-ray flux sensitivity degrades, while the flux sensitivity of neutrino detectors does not change significantly. Presently, this leaves room for the discovery of extended neutrino sources in our Galaxy,  given also the fact that the Galactic Disk has not been homogeneously covered by the current IACT arrays. On the other hand, in-depth surveys of the Galactic Disk in coming years by CTA  could significantly improve this situation. Alternatively,  in a more optimistic scenario, CTA could reveal on the sky bright extended regions of multi-TeV gamma-rays, and thus indicate the sites of potential detectable sources of VHE neutrinos. Here, we study this question based on the comparative analysis of the sensitivities of CTA and KM3NeT for extended sources. For this purpose, we have developed a common approach for calculations of sensitivities of CTA and KM3NeT. The method is based on the analytical parametrization of the main quantities (as functions of energy) characterizing the process of detection of gamma rays and neutrinos: the effective detection area,  the point spread function, the energy resolution and the background rates. These functions  for CTA have been provided in our previous work \cite{Ambrogi:2016hmm} using the results of the publicly available simulations performed by the CTA  consortium. In this paper we present similar parametrizations for the neutrino detector based on the simulation results published by the KM3NeT collaboration \cite{Adrian-Martinez:2016fdl}. Similar results are expected to hold for the IceCube-Gen2 detector \cite{icgen2}, whose performances however are not yet publicly available. Hence, we will focus on KM3NeT only here. \\
The paper is structured as follows: in Sec.~\ref{sec:detectors} we discuss the performances of CTA and KM3NeT, in particular their angular resolution, effective area and expected background rates. Then, in Sec.~\ref{sec:sens} we describe the procedure defined to compute the instrument sensitivity, considering different sizes of the sources and analyzing the different impact they have on the sensitivity of these instruments. As an application of this study, in Sec.~\ref{sec:sources} we consider the case of two galactic objects, for which the gamma-ray and neutrino connection has been widely discussed in literature \cite{Kappes:2006fg, Vissani:2006tf, Aharonian:2008zza, Vissani:2011ea}. The young supernova remnant (SNR) RX J1713-3946 is presented in Sec.~\ref{sec:rxj}, while the region of the Galactic Center Ridge is investigated in Sec.~\ref{sec:gcr}, being both realistic candidate neutrino sources \cite{Adrian-Martinez:2016fei, Costantini:2004ap, AlvarezMuniz:2002tn, Abramowski:2016mir, Celli:2016}. In addition to these scenarios, the second HAWC catalog of TeV sources is considered in Sec.~\ref{sec:2hwc}, where potential sources for a neutrino detection are highlighted. Eventually, conclusions are derived in Sec.~\ref{sec:disc}.  \\

In this study, we do not explore different possible improvements of the sensitivities of both detectors by applying dedicated tools for the background rejection and for the reconstruction of the gamma-ray and neutrino induced events from extended sources. Further details on these dedicated tools are given in Sec.~\ref{sec:sens}. Therefore, we cannot exclude some deviations of our results from the upcoming, more detailed and sophisticated studies by the CTA and KM3NeT consortia. In this regard, the results presented in  this work can be considered as conservative estimates of sensitivities of both CTA and KM3NeT.

%%%%%%%%%%%%%%%%%%%%%%%%%%%%%%%%%%%%%%%%%%%''''
\section{Detector performances}
\label{sec:detectors}
%%%%%%%%%%%%%%%%%%%%%%%%%%%%%%%%%%%%%%%%%%%''''
In this Section we discuss the sensitivities of CTA and KM3NeT. Both instruments are based on the Cherenkov technique, detecting the light induced by the passage of an ultra-relativistic charged particle in a given medium: the air in the case of IACTs and the water or the ice in the case of neutrino detectors. 
Although the same physical principle is applied, the reconstruction of the signal parameters and the background rejection are quite different. Both telescopes operate in the TeV domain, reaching the best performance between 1 and 10~TeV in the case of CTA and 10 to 100 TeV in the case of KM3NeT.

\subsection{The Cherenkov Telescope Array}
\begin{figure*}[!h]
  \centering
  \begin{tabular}[b]{c c}
    \includegraphics[width=0.5\textwidth]{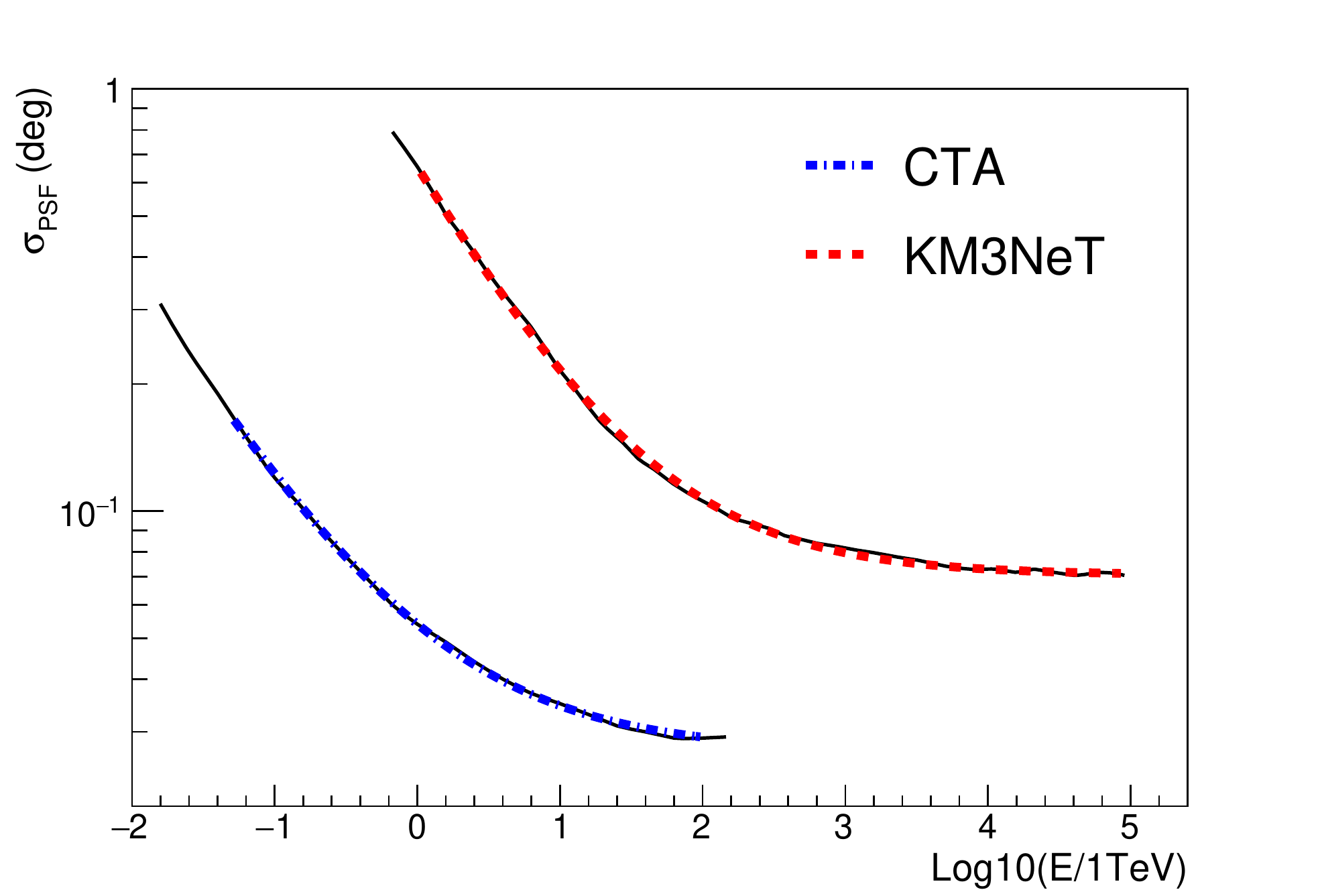} & \includegraphics[width=0.5\textwidth]{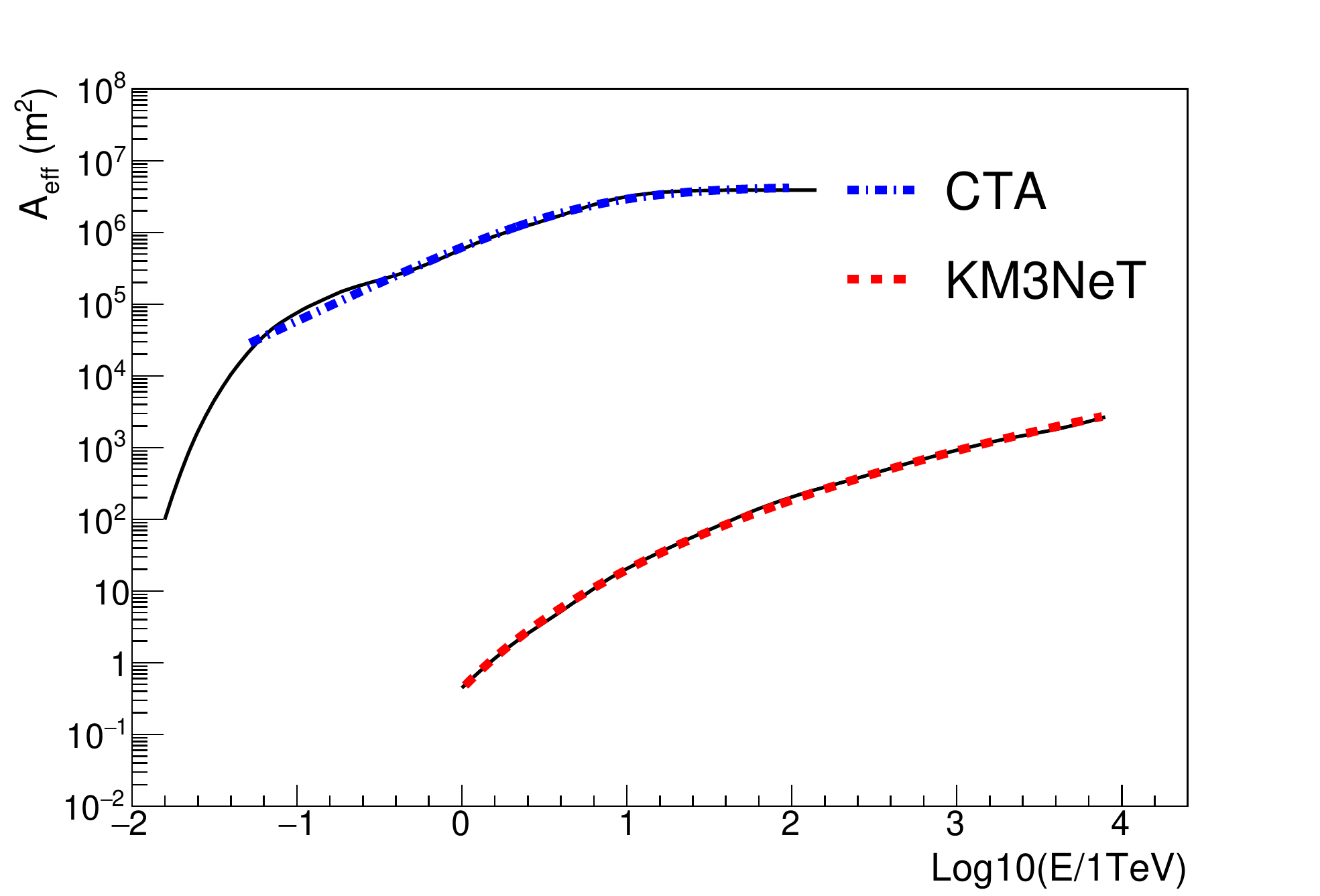} \\
    \multicolumn{2}{c}{\includegraphics[width=0.5\textwidth]{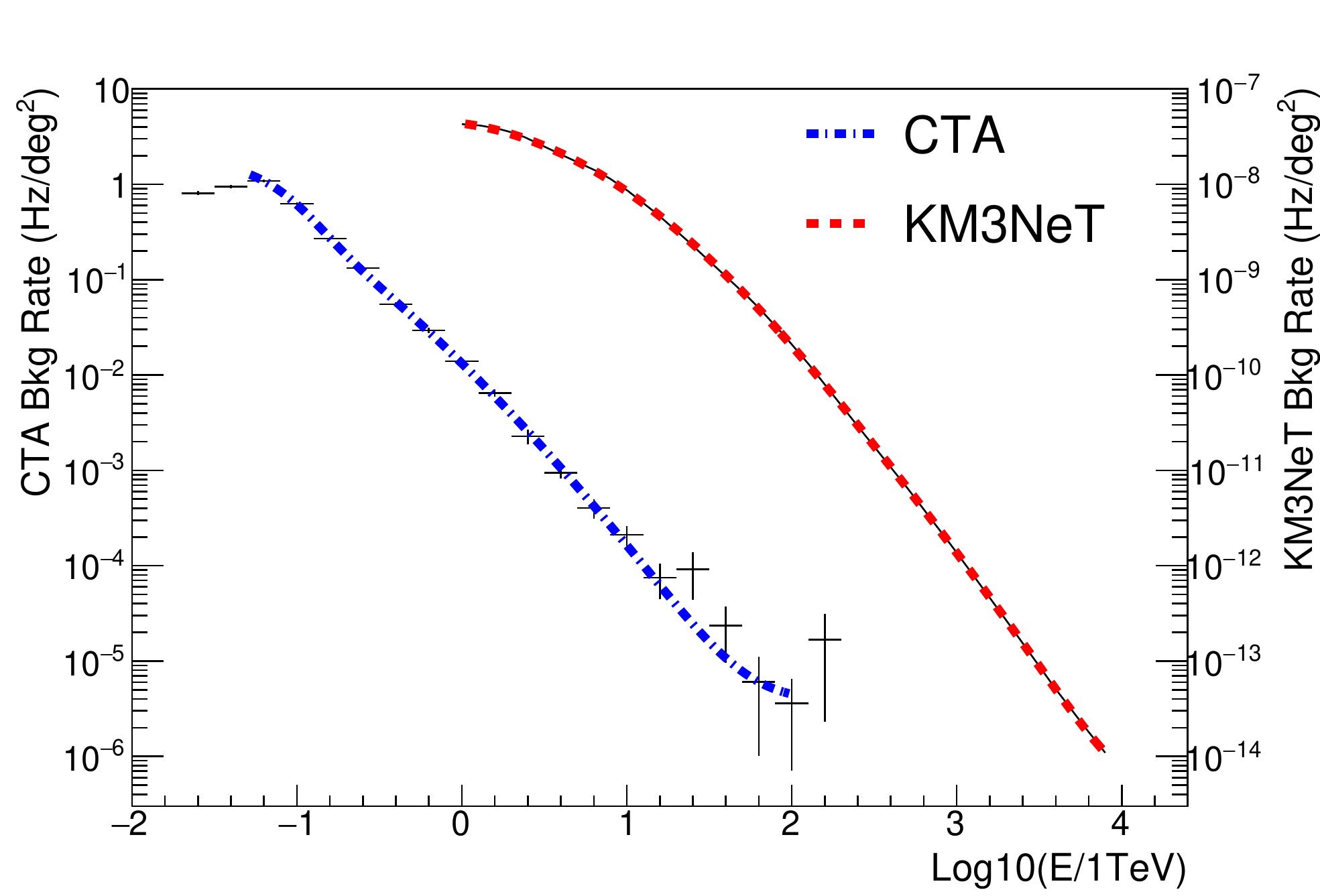} } \\
  \end{tabular} \qquad
  \caption{Energy dependent performance of the CTA and KM3NeT telescopes, i.e. their angular resolution (top-left), effective area (top-right) and background rates per unit of solid angle (bottom). For both the two telescopes, the black lines corresponds to the publicly available instrument responses, respectively for CTA Southern array \cite{CTAperformaceWeb} and KM3NeT \cite{Adrian-Martinez:2016fdl}. Dotted curves are the best fit, valid in the energy range $E\in[0.05-100]$~TeV for CTA and above 1~TeV for KM3NeT. The corresponding analytical parameterizations are reported in Tab.~\ref{tab:irf} and in Tab.~\ref{tab:irfKM3} for CTA and KM3NeT, respectively. The KM3NeT angular resolution is for $\nu_\mu$ charged current events and the muon neutrino effective area (six blocks) corresponds to triggered events with a zenith angle greater than $80^{\circ}$, averaged over both $\nu_\mu$ and $\bar{\nu}_\mu$. The same trigger conditions are exploited for the background rate computation. The atmospheric muon neutrino background here considered accounts for the conventional component from light meson decay \cite{Honda:2016qyr} and for contribution from heavy hadrons \cite{Enberg:2008te}. For the details on the CTA IRFs and their parameterizations, the reader is referred to \cite{Ambrogi:2016hmm}.}
\label{fig:performance}
\end{figure*}

Although the principle of detection of gamma-rays by CTA is almost identical to the operation of the current  H.E.S.S., MAGIC and VERITAS stereoscopic systems of IACTs,  the angular resolution of CTA will be reduced down to 1-2 arcminutes, and the flux sensitivity will be improved compared its predecessors, by  one order of magnitude. 
In order to view the whole sky, CTA will consist of two arrays of IACTs, one in the Northern (La Palma, Canary Islands) and one in the Southern (Paranal, Chile) hemisphere. The Southern array is aimed to study the major fraction of the Galactic plane including the Galactic Center region. 
One of the proposed layouts for the Southern observatory, the so-called \emph{2-Q} layout, consists of 4 large size telescopes (LSTs; 23 meter class, field of view (FoV) of the order of 4.5~deg) optimized for detections below 100~GeV, 24 medium size telescopes (MSTs; 12 meter class, FoV of 7~deg), covering the core energy of CTA, i.e. 100~GeV to 10~TeV, and 72 small size telescopes (SSTs; 4 meter class, FoV ranging from 9.1 to 9.6~deg), sensitive to energies above 10~TeV \cite{Hassan:2015bwa}. For this configuration, publicly available instrument response functions (IRFs) have been released by the CTA Consortium\footnote{The publicly available CTA performance files can be accessed at \cite{CTAperformaceWeb}.}, obtained through detailed Monte Carlo (MC) simulations of a point-like object placed at the center of the FoV and observed at a zenith angle of 20~deg (averaged between north/south-wise in azimuth). In our previous work \cite{Ambrogi:2016hmm} we parametrized these IRFs by simple analytical functions of energy. The results are presented in Tab.~\ref{tab:irf} and Fig.~\ref{fig:performance}.  
It should be noted that the IRFs released by the CTA Collaboration, and here considered, are the derived best responses which maximize at each energy bin the CTA differential sensitivity to point-like sources. Therefore, an improvement of the instrument performance is expected when using analysis cuts aimed to maximize the telescope potential to extended objects, which are the main topic of this paper.

\subsection{The KM3 Neutrino Telescope}
High-energy neutrino telescopes are three dimensional arrays of photomultipliers, where Cherenkov radiation produced by the neutrino interaction products are observed: the position, time and charge deposit are used to infer both the direction and the energy of the incoming neutrino. KM3NeT is an under-construction neutrino telescope \cite{Biagi2017}, located deep in the Mediterranean sea. In its final configuration, the detector will consist of 6 building blocks, each instrumented with 115 vertical detection units (DU): a DU is composed of 18 digital optical modules, each containing 31 PMTs \cite{Adrian-Martinez:2016fdl}. Although neutrino detectors are sensitive to all neutrino flavors, from the point of view of reconstruction of arrival directions of primary neutrinos, the best channel is represented by charged current interactions of muon neutrinos resulting in the production of a muon, which is experimentally visible as a track. 
The main background source for this event sample is then constituted by atmospheric muons and the accompanying atmospheric neutrinos. In order to reduce atmospheric muons, only events coming from below the detector horizon are selected (the so called upward going sample), as they are absorbed in their path through the Earth. Although in this way the visibility to a given source is decreased, a cleaner sample of events can be selected. Alternatively, also events coming from above the detector horizon can be used, selecting a reduced fiducial volume of the detector. This is possible for high-energy events: in the case of KM3NeT, preliminary studies presented in \cite{Adrian-Martinez:2016fdl} show that the containment requirements on muons with $E_{\mu}>10$~TeV reduced the volume by about 20\%. The same containment technique allows to include cascade events to neutrino searches: the usage of this additional channel would allow to significantly increase the sample statistics, given the full sky coverage of this event sample, lowering the instrument sensitivity, as extensively discussed in \cite{Adrian-Martinez:2016fdl}. 

\begin{table*}[!h]
\caption{\label{tab:irf}Energy dependent analytical parameterizations for the angular resolution ($\sigma_{PSF}$), the effective area ($A_{eff}$) and the background rate per solid angle after the rejection cuts ($\textrm{BgRate}$) of CTA, as reported in \cite{Ambrogi:2016hmm}. These formula have been obtained as best-fit to the publicly available IRFs for a possible layout of the CTA Southern array \cite{CTAperformaceWeb}. The energy range of validity is from 50~GeV to 100~TeV. For the details of the IRFs as obtained by the CTA Collaboration, the reader is referred to \cite{Bernlohr:2012we, Hassan:2015bwa}.}
\begin{center}
\begin{tabular}{  l  l  }
  \hline  
  							&	CTA response parameterization, with $x=\log_{10}(E/1~\mbox{TeV})$   \\
  \hline
  $ $ 							&	  $ $ \\  
  $\sigma_{PSF}~[\mbox{deg}]$ 	& 	  $\sigma_{PSF} (x) = A \cdot \left[ 1 + \exp \left( - \dfrac{x}{B} \right) \right]$ \\
  $ $ 							&	  $ $ \\  
  $ $ 							&	 $A=2.71\cdot10^{-2}$~deg  ~~~~ $B=7.90\cdot10^{-1}$ \\
  $ $ 							&	  $ $ \\  
  \hline  
  $ $ 							&	  $ $ \\  
  $A_{eff}~[\mbox{m}^2]$			& 	  $A_{eff}(x) = A \cdot \left[1 + B \cdot \exp \left( -\dfrac{x} {C} \right) \right]^{-1}$ \\
  $ $ 							&	  $ $ \\
  $ $ 							&	  $A=4.36 \cdot 10^6$~m$^2$ ~~~~ $B=6.05$ ~~~~ $C=3.99\cdot10^{-1}$ \\
  $ $ 							&	  $ $ \\  
  \hline  
  $ $ 							&	  $ $ \\  
  $\textrm{BgRate}~[\mbox{Hz/deg}^2]$		& 	  $\textrm{BgRate}(x) = A_1 \cdot \exp \left( - \dfrac{(x-\mu_1)^2}{2\cdot\sigma_1^2}\right) + A_2 \cdot \exp \left( - \dfrac{(x-\mu_2)^2}{2\sigma_2^2}\right) + C $\\
  $ $ 							&	  $ $ \\  
  $ $							& 	  $A_1=3.87\cdot10^{-1}$~Hz/deg$^2$ ~~~~ $\mu_1=-1.25$ ~~~~ $\sigma_1=2.26\cdot10^{-1}$ \\
  $ $							& 	  $A_2=27.4$~Hz/deg$^2$ ~~~~  $\mu_2=-3.90$ ~~~~ $\sigma_2=9.98\cdot10^{-1}$ ~~~~ $C=3.78\cdot10^{-6}$~Hz/deg$^2$ \\
  $ $ 							&	  $ $ \\  
  \hline  
\end{tabular}
\end{center}

\caption{\label{tab:irfKM3}Energy dependent analytical parameterizations for the angular resolution ($\sigma_{PSF}$), the effective area ($A_{eff}$)  and the background rate per solid angle ($\textrm{BgRate}$) to track-like events for the six building block configuration of KM3NeT. The effective area and background refer to the event selection after the cut on the zenith angle: only reconstructed up-going muons are considered. See \cite{Adrian-Martinez:2016fdl} for the details. The energy range of validity is $E_{\nu} \ge 1$~TeV.}
\begin{center}
\begin{tabular}{  l  l  }
  \hline  
  							&	KM3NeT response parameterization, with $x=\log_{10}(E/1~\mbox{TeV})$   \\
  \hline
  $ $ 							&	  $ $ \\  
  $\sigma_{PSF}~[\mbox{deg}]$ 	& 	  $\sigma_{PSF} (x) = A \cdot  \exp \left( - \dfrac{x}{B} \right) + C$ \\
  $ $ 							&	  $ $ \\  
  $ $ 							&	 $A= 5.88\cdot10^{-1} $~deg  ~~~~ $B=7.19\cdot10^{-1}$ ~~~~ $C=6.95\cdot10^{-2}$~deg \\
  $ $ 							&	  $ $ \\       
  \hline  
  $ $ 							&	  $ $ \\  
  $A_{eff}~[\mbox{m}^2]$			& 	  $A_{eff}(x) = A  \cdot \left(1 + x \right)^{B}$ \\
  $ $ 							&	  $ $ \\
  %$ $ 							&	  $A=0.14$~m$^2$ ~~~~ $B=5.51$ \\ % with zenith cut 2bb 1.44638e-01, 5.51398e+00 
  $ $ 							&	  $A=0.43$~m$^2$ ~~~~ $B=5.51$ \\ % with zenith cut 6bb, 4.33915e-01, 5.51397e+00
  $ $ 							&	  $ $ \\  
  \hline  
  $ $ 							&	  $ $ \\      
  $\textrm{BgRate}~[\mbox{Hz/deg}^2]$		& 	  $\textrm{BgRate}(x) = A_1 \cdot \exp \left( - \dfrac{(x-\mu_1)^2}{2\cdot\sigma_1^2}\right) + A_2 \cdot \exp \left( - \dfrac{(x-\mu_2)^2}{2\sigma_2^2}\right) + C $\\
  $ $ 							&	  $ $ \\  
%  $ $							& 	  $A_1=2.95\cdot10^{-11}$~Hz/deg$^2$ ~~~~ $\mu_1=9.80\cdot10^{-1}$ ~~~~ $\sigma_1=6.70\cdot10^{-1}$ \\
%  $ $							& 	  $A_2=1.57\cdot10^{-8}$~Hz/deg$^2$ ~~~~  $\mu_2=-2.58\cdot10^{-1}$ ~~~~ $\sigma_2=6.75\cdot10^{-1}$ ~~~~ $C=1.31\cdot10^{-15}$~Hz/deg$^2$ \\ %  with zenith cut, 2bb
  $ $							& 	  $A_1=6.76\cdot10^{-10}$~Hz/deg$^2$ ~~~~ $\mu_1=2.89\cdot10^{-1}$ ~~~~ $\sigma_1=7.55\cdot10^{-1}$ \\
  $ $							& 	  $A_2=4.58\cdot10^{-8}$~Hz/deg$^2$ ~~~~  $\mu_2=-2.37\cdot10^{-1}$ ~~~~ $\sigma_2=6.61\cdot10^{-1}$ ~~~~ $C=3.53\cdot10^{-15}$~Hz/deg$^2$ \\ % with zenith cut, 6bb
  $ $ 							&	  $ $ \\    
  \hline  
\end{tabular}
\end{center}
\end{table*} 

The IceCube detector \cite{ICfirst} has validated the search strategy for the detection of a neutrino signal: the first evidence for an extra-terrestrial flux of neutrinos was reported in \cite{ICdiscovery}. Still nowadays, the origin of such a signal is unknown: it was pointed out in \cite{PalladinoVissani2016} and \cite{Neronov:2015} that part of the detected flux might have been originated in the Galactic Plane. However, the latest constraints from both the ANTARES and the IceCube telescopes strongly tighten the possible Galactic contribution to the measured diffuse neutrino flux, as reported in \cite{2017antaresGP,2017icGP}. Further investigation of this scenario is necessary: a more extended statistical sample is required to test different scenarios, which is going to be provided by the next generation of neutrino telescopes. A good angular resolution is necessary to identify sources: water-based telescopes, as KM3NeT, will be able to reach an angular resolution as low as $0.2^{\circ}$ at 10 TeV in the track channel, as shown in Fig.~\ref{fig:performance} (top-left). However, for very extended sources (more than $2^{\circ}$ in radius), also cascade events could be used in principle for astronomy: the atmospheric background for shower events is significantly smaller than that for muon neutrinos, allowing for a clearer signal detection. Finally, the energy resolution is a very important goal as well: in neutrino telescopes, the energy of the muon is reconstructed through the energy deposited in the detector, therefore it is only a lower limit to the true neutrino energy. The energy resolution obtained for muon events fully contained in the detector is 0.27 units in $\log_{10}(E_\mu)$ for 10 TeV $\leq E_{\mu} \leq$ 10 PeV \cite{Adrian-Martinez:2016fdl}. The case of cascade events provides a better energy resolution, given that they develop entirely very close to the interaction point. 

The effective area of the detector to up-going events is given in Fig.~\ref{fig:performance} (top-right): it refers to the 6 building blocks configuration of the KM3NeT detector. This will correspond to an effective area of $\sim 1000$~m$^2$ at high energies, where the long muon range extends the volume within which neutrino interactions can be detected. An analytical representation of this effective area is given in Tab.~\ref{tab:irfKM3}, valid for $E_{\nu}\ge1$~TeV; similarly, an analytical representation of the track angular resolution is given in the same table. Some words of caution are mandatory here: the effective area strongly depends on the source position and on the background conditions, which affect in a crucial way the selection of events. Moreover, optimized selection is usually dependent on the specific analysis: in order to properly evaluate the detector performances, detailed simulations of such features are necessary, which are performed by the Collaboration itself. Such a tailored selection might result into a relevant improvement of the instrument sensitivity. The effective area used in the following refers to triggered events, reconstructed with a zenith angle greater than $80^{\circ}$, as presented in \cite{Adrian-Martinez:2016fdl}. 

For the sensitivity estimation we adopt the conservative approach of considering sources below the horizon, since this is the cleanest procedure to identify a signal, providing a high suppression of the background atmospheric muon flux. The remaining atmospheric neutrino flux is composed of two contributions: a \emph{conventional} component, due to the decay of light mesons from atmospheric air showers, and a \emph{prompt} component due to the decay of charmed hadrons. The atmospheric neutrino background is evaluated following \cite{Adrian-Martinez:2016fdl}, with the conventional model from \cite{Honda:2016qyr} and the prompt model from \cite{Enberg:2008te}, which becomes dominant over the conventional flux at $E_{\nu}>1$~PeV: the expected up-going neutrino background rate is shown in Fig.~\ref{fig:performance} (bottom) while the corresponding analytical parameterization is reported in Tab.~\ref{tab:irfKM3}.  

%%%%%%%%%%%%%%%%%%%%%%%%%%%%%%%%%%%%%%%%%%%''''
\section{Sensitivity to extended sources}
\label{sec:sens}
%%%%%%%%%%%%%%%%%%%%%%%%%%%%%%%%%%%%%%%%%%%''''

A common procedure for gamma-ray and neutrino telescopes is here introduced for the computation of the sensitivity curves: the same analytical approach is applied to the two detectors to calculate their sensitivities to facilitate their comparison. 
For calculations of the minimum detectable fluxes of gamma-rays and neutrinos we use the relevant functions from Tab.~\ref{tab:irf} and Tab.~\ref{tab:irfKM3}.
Note that the curves shown in Fig.~\ref{fig:sens}-\ref{fig:hawc_2} correspond to the differential sensitivities: the per bin sensitivity allows not only the identification of a source but also its spectroscopic analysis. Since the sensitivity curves have to be compared point by point, we use the same energy binning for both the gamma-ray and the neutrino sensitivities. We define three bins per logarithmic decade, so that the energy resolution of both instruments is covered in each bin. \\
The minimum detectable flux is defined as the flux that gives in each energy bin:
\begin{itemize}
\item[1)] a minimum number of signal events, $N_s^{min}$;
\item[2)] a minimum significance level of background rejection, $\sigma_{min}$;
\item[3)] a minimum signal excess over the background uncertainty level.
\end{itemize}
Therefore, the instrument sensitivity is fixed by the one condition among the three listed above which dominates over the other two. The number of signal events, $N_s$, is obtained through the folding of an $E^{-2}$ power-law spectrum with the instrument response. We set $N_s^{min}=10$ for CTA and $N_s^{min}=1$ for KM3NeT. 
The significance level of the detection is expressed by the standard deviation $\sigma$, defined as: 
\begin{equation}
\sigma= \frac{N_s}{\sqrt{N_b}}
\end{equation}
where $N_b$ is the number of background events in the energy bin. The threshold on the minimum number of $\sigma$ is set to $\sigma_{min}=5$ for the gamma-ray telescope and to $\sigma_{min}=3$ for the neutrino telescope. The values of $N_s^{min}$ and $\sigma_{min}$ are reduced in the case of neutrino telescopes in order to investigate the limits of the source detection capability, given that neutrino astronomy is not properly yet at its dawn. 
However, we have to keep in mind that each energy bin is satisfying all the above criteria: therefore, for instance, a differential $3\sigma$ requirement corresponds to an actual higher significance in the energy bins where this is not the dominant condition. \\
Concerning the condition on the background uncertainty, in the case of CTA we assume a 1\% systematic uncertainty on the modeling of the background and require a signal of at least five times this background accuracy level, i.e. $N_s/N_b \ge 0.05$, following the approach adopted by CTA \cite{Bernlohr:2012we}. In the case of neutrinos, instead, we follow the approach adopted by KM3NeT in \cite{Adrian-Martinez:2016fdl} and assume a 25\% background systematic uncertainty, mainly related to uncertainty in the theoretical modeling of the atmospheric neutrino background. This uncertainty might be reduced in the future, adopting a data-based evaluation of background fluctuations. Therefore, in the neutrino case, a signal of at least three times higher than the background accuracy is assumed: this converts into requiring a signal to background ratio of at least $N_s/N_b \ge 0.75$. Moreover, to account for the statistical fluctuations of the background, the number $N_b$ is randomized according to a Poissonian distribution, with the results being averaged over 1000 realizations of the sensitivity estimation. An observation time of 50 hours is assumed for the gamma-ray telescope, while 10 years are assumed for the neutrino telescope. \\

Note that the sensitivities are calculated without optimization of the tools for reconstruction of characteristics of primary particles and without exploring different dedicated background-rejections methods. A precise estimation of the CTA capabilities for the detection of extended sources would require a complete 3D analysis and the study of sub-structures on arcminute scales, as the arcminute PSF of CTA would permit to resolve the morphological details of many extended sources beyond the disk-like structure here considered. To our knowledge, such studies are currently being conducted by the CTA consortium. However, morphological studies are not an easy task for neutrino astronomy, because of the fainter flux when only a reduced part of the source is considered. As from the neutrino side, tailored background rejection techniques targeted to the suppression of atmospheric muons and neutrinos might be included at the analysis level: for instance, vetoing few external layers of the detector has been demonstrated to be effective in rejecting background muons \cite{ICdiscovery}. Furthermore, the same technique constitutes a powerful method to reject also downward-going atmospheric neutrinos, through the identification of the accompanying muon, as proposed in \cite{resconi}. However, since we here propose a comparison between gamma-ray and neutrino telescopes, throughout this work we limit our estimation of the instrument potentials to the analysis technique previously described.

\subsection{Source angular extension}
The radial dimension of the extended source strongly affects the sensitivity for their detection. We consider here eight different source sizes, i.e. $R_{src}=[0.1,0.2,0.5,0.8,1.0,1.2,1.5,2.0]$~deg. The largest size has been fixed as a conservative threshold value for which the degradation of the CTA response with the off-axis angle does not play a significant effect and therefore the IRFs meant for sources located close to the center of the FoV are still valid (see e.g. \cite{Szanecki:2015zaa} for a study on the expected CTA off-axis performance). Publicly available IRFs for objets placed off-axis have not been released by the CTA consortium yet. However, an estimation of the worsening of the CTA sensitivity due to the off-axis pointing of point-like sources is presented in \cite{CTAperformaceWeb}. On the basis of this result, we evaluate a correction factor to the flux sensitivity for objects with an extension of up to $2$~deg. This is presented in \ref{sec:offaxis}, in the energy bins where the CTA results are available: we show that the sensitivity worsening is always lower than a factor of two, even for sources as large as $2$~deg. \\
In the sensitivity computation, the angular resolution $\sigma_{PSF}$ affects the actual size of the observed region of interest (ROI). We define the radius of the ROI as:
\begin{equation}
\label{eq:rroi}
R_{ROI}=\sqrt{\sigma^2_{PSF}+R^2_{src}}
\end{equation}
We consider spherical sources placed at the center of the FoV, covering a solid angle $\Omega=\pi R_{ROI}^2$ for the background computation. The resulting sensitivity curves are shown in Fig.~\ref{fig:sensCTA} for the gamma-ray telescope and in Fig.~\ref{fig:sensKM3} for the neutrino detector. Previous works on the potential of current neutrino telescopes to observe extended sources \cite{2014antaresExtSrc,2014icExtSrc} show that the next-generation neutrino telescope will push sensitivity limits down by more than an order of magnitude with respect to current instrument sensitivities. \\

\subsection{Discussion on sensitivity curves}
\begin{figure*}[htb]
\centering
\subfigure[]{\label{fig:sensCTA}\includegraphics[width=0.49\textwidth, height=0.27\textheight]{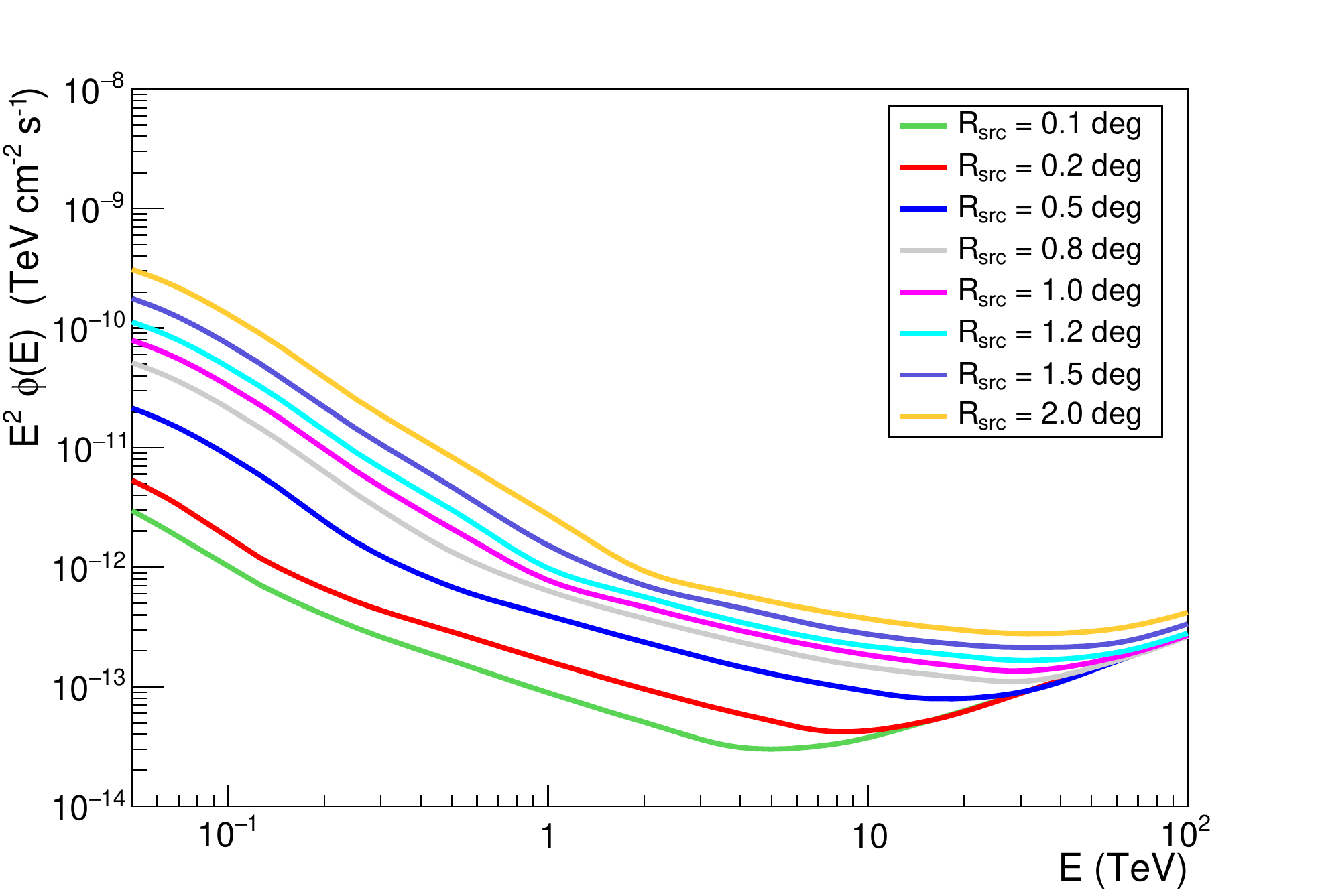}}
\subfigure[]{\label{fig:sensKM3}\includegraphics[width=0.49\textwidth, height=0.27\textheight]{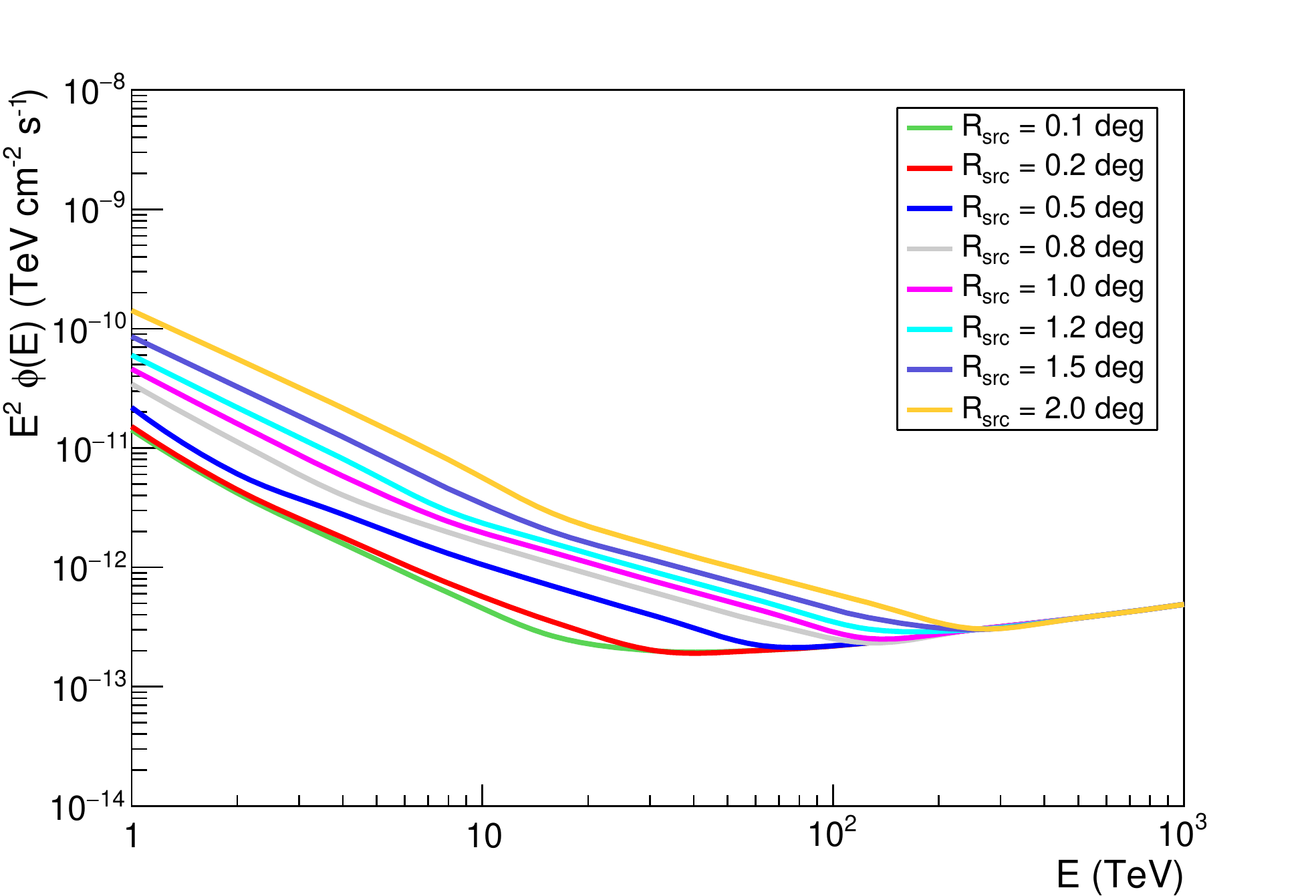}}
\caption{Minimum detectable flux computed according to the procedure described in Sec.~\ref{sec:sens} in the case of extended sources for: (a) CTA (50 hours observation time) and for (b) KM3NeT (10 years exposure time).}
\label{fig:sens}
\end{figure*}

Fig.~\ref{fig:sensCTA} and Fig.~\ref{fig:sensKM3} demonstrate that the deterioration of the sensitivity with source size shows an energy dependence for both instruments. In principle, a simple re-scaling of the point-source sensitivity according to the actual extension of the source (i.e. through an energy-dependent scaling-factor proportional to $R_{ROI}/\sigma_{PSF}$), would predict a stronger deterioration of the sensitivity for extended objects at energies at which the angular resolution is smaller. Thus, since the angular resolution is improved with energy, one would expect a stronger effect at higher energies. Nevertheless, Fig.~\ref{fig:sens} does not show such a tendency for both telescopes. The reason being that at very high energies the detection of the signal proceeds at the very low background rate, thus the detection condition is determined by the signal statistics rather than by the background, i.e. by condition 1) listed in Sec.~\ref{sec:sens}. Indeed, it is seen from Fig.~\ref{fig:sens} that the sensitivities become almost independent of the source extension at a few tens of TeV for both gamma-rays and neutrinos.\\
According to the same arguments, one would expect a reduced dependence on the object size at low energies, where the angular resolution worsens. Nevertheless, at these energies, we find the maximum deviation between different sizes. This is because at lower energies, the sensitivity depends on the background as $N_s / \sqrt{N_b} \propto N_s / R_{ROI}$ when the condition 2) holds (at intermediate energies), while it follows a $N_s / N_b \propto N_s / R_{ROI}^2$ dependence when the background systematics are mainly affecting the signal identification, such that condition 3) dominates (at the lowest energies). Consequently, the increase of degradation with source size is maximized at the lowest energy, as the sensitivity deteriorates as $\propto R_{ROI}^2$. This is especially evident in the case of CTA, while a less significant degradation with the source size takes play in the case of KM3NeT. The reason for this difference lies in the fact that, contrary to CTA, the ROI that defines the KM3NeT angular search window at the lowest energies is dominated by the size of instrument PSF.  Consequently, the KM3NeT angular resolution affects the minimum detectable flux as long as $R_{src} < \sigma_{PSF}$, with $\sigma_{PSF}\simeq0.7$~deg around 1~TeV. \\
In summary, we see a deviation with increasing extension of the source size which is maximum at low energies, reduces towards higher energies, and eventually disappears at the highest energies.

%%%%%%%%%%%%%%%%%%%%%%%%%%%%%%%%%%%%%%%%%%%''''
\section{The case of RX J1713.7-3946 and the Galactic Center Ridge}
\label{sec:sources}
%%%%%%%%%%%%%%%%%%%%%%%%%%%%%%%%%%%%%%%%%%%''''
As the origin of the cosmic-ray flux measured on Earth is still a matter of debate, it is mandatory to investigate sources which might be responsible for it. Galactic sources are believed to contribute up to energies of about 1~PeV, where the so called \textit{knee} is located. Among them, young SNRs represent promising candidates, given that the strong shocks produced during the supernova explosion might be able to accelerate particles, as predicted in Diffusive Shock Acceleration (DSA) scenarios \cite{bell1978,blandford78}. However, the lack of observational evidence of PeV protons from such objects doesn't permit yet to firmly establish the SNR paradigm for the origin of Galactic cosmic rays, and future gamma-ray and neutrino observations are needed to further constrain theoretical models. In this regard, the recent claim of a PeVatron in the center of our own Galaxy \cite{Abramowski:2016mir} opens a new possibility to explain the flux of cosmic rays below the \textit{knee} and deserves a deeper investigation, both from the gamma-ray and the neutrino side. For these reasons, in this Section we discuss the case of two bright extended gamma-ray sources: the young SNR RX J1713.7-3946 and the Galactic Center Ridge. 
In the following, the muon neutrino fluxes expected from these sources are computed according to the model in \cite{Villante:2008qg}, assuming a 100\% hadronic origin of the measured gamma-ray flux and no internal absorption. Both the measured gamma-ray and expected neutrino fluxes are shown together with the detector sensitivity curves. Fluxes are reported in a binned form, such that it is directly possible to compare the expected source flux in a given energy range with the detector sensitivity in the same band. These binned fluxes are defined as:
\begin{equation*}
\phi(E) = \frac{1}{\Delta E} \int_{E_{\textrm{min}}}^{E_{\textrm{max}}} \frac{dN}{dE} dE
\end{equation*}
where $\Delta E=E_{\textrm{max}}-E_{\textrm{min}}$ represents the amplitude of the logarithmic bins used in the sensitivity computation. The fluxes are reported together with the associated error bands. In the case of gamma rays, we consider the statistical errors of the estimated spectral parameters and then, for each energy bin, we compute the upper/lower band as the curve defined by the combination of parameters ($\pm$ statistical errors) for which the flux is maximized/minimized. The same approach is adopted also for the neutrino flux error bands: a scanning of the neutrino fluxes resulting from all the different combinations of gamma-ray parameters is performed and the maximum/minimum neutrino flux computed.

%%%%%%%%%%%%%%%%%%%%%%%%%%%%%%%%%%%%%%%%%%%''''
\subsection{RX J1713.7-3946}
\label{sec:rxj}
\begin{figure*}[ht]
\centering
\subfigure[]{\label{fig:RXJ} \includegraphics[width=0.49\textwidth]{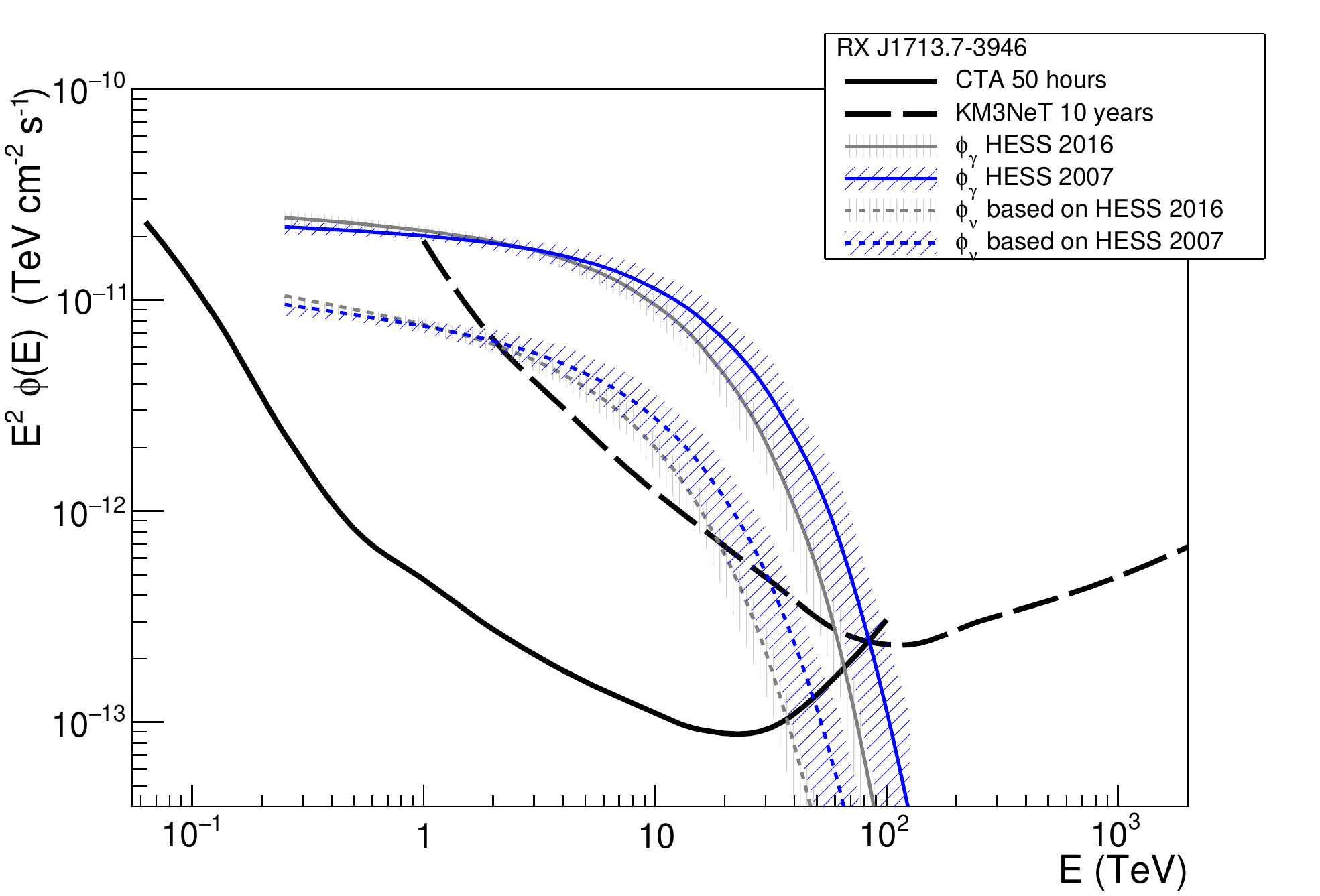}}
\subfigure[]{\label{fig:GCR} \includegraphics[width=0.49\textwidth]{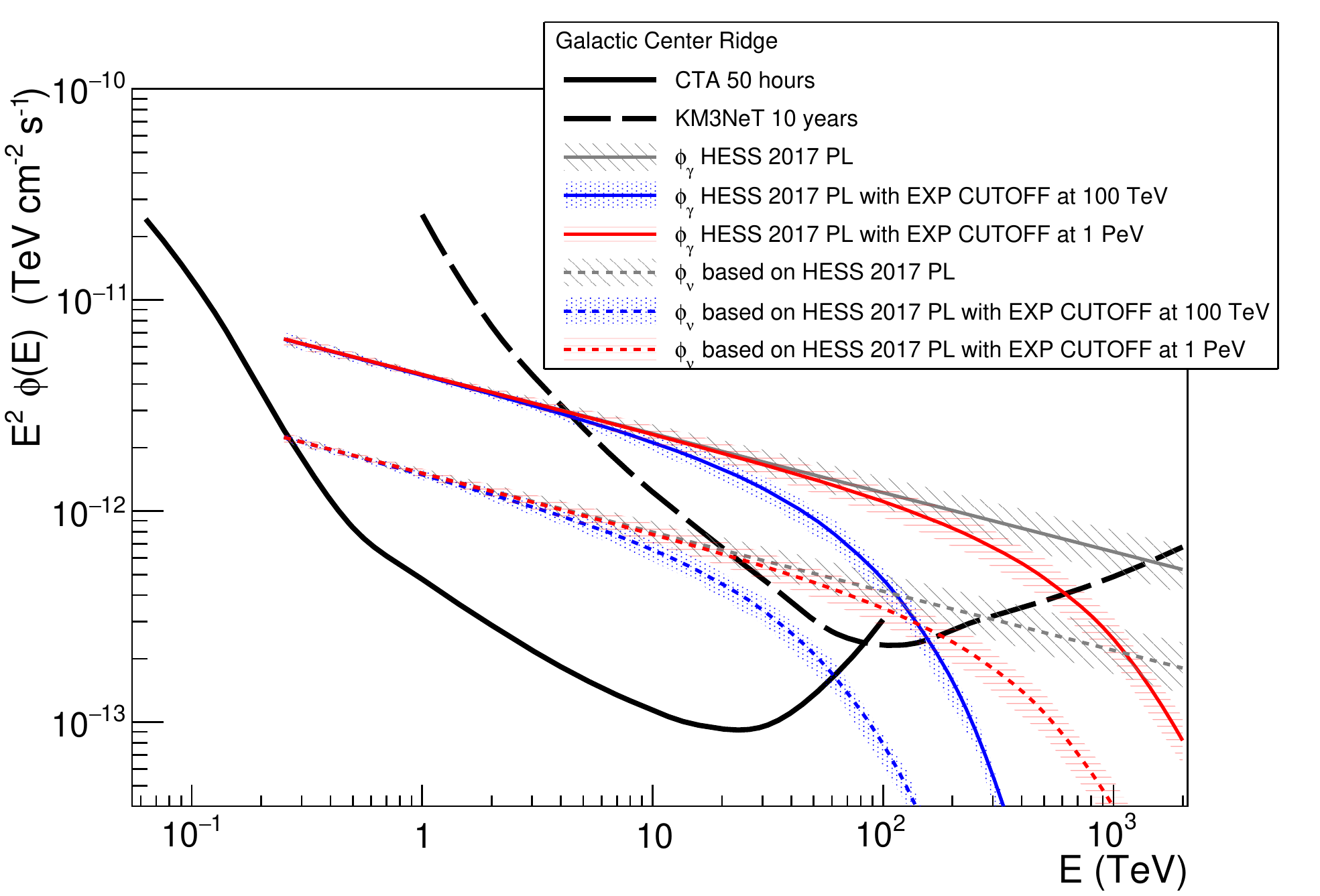}}
\caption{Minimum detectable flux computed according to the procedure described in Sec.~\ref{sec:sens} for CTA and KM3NeT to: (a) the extended SNR RX J1713.7-3946 (spherical source with radius of $0.6$~deg) and (b) the diffuse emission from the Galactic Center Ridge (rectangular box with longitudinal size of 2.0~deg and latitudinal size of 0.6~deg). The binned gamma-ray fluxes are shown as colored solid lines. The dashed curves are the binned muon neutrino fluxes computed according to the model in \cite{Villante:2008qg}.}
\end{figure*}

%%%%%%%%%%%%%%%%%%%%%%%%%%%%%%%%%%%%%%%%%%%''''
The case of this source is of great interest for neutrino telescopes, given that it is the brightest SNR in the TeV sky. 
Moreover, its location in the sky makes it observable with up-going events at the latitude of KM3NeT for 70\% of the time. The recent data from the H.E.S.S. Collaboration  \cite{Abdalla:2016vgl} suggest a spectrum in the form of a power law with exponential cutoff:
\begin{equation}
\frac{dN_{\gamma}}{dE}(E)  = \phi_0 \left( \frac{E}{E_0} \right)^{-\alpha} \exp{ \left[ - \frac{E}{E_{cut}} \right]}
\end{equation}
with $E_0=1$~TeV, $\phi_0=2.3\times 10^{-11}$~TeV$^{-1}$~cm$^{-2}$~s$^{-1}$, $\alpha=2.06$ and $E_{cut}=12.9$~TeV for the best-fit model. Note that the best-fit flux of the source, published in the previous paper \cite{Aharonian:2006ws} of the H.E.S.S. collaboration, with the parameters $\phi_0=2.13\times 10^{-11}$~TeV$^{-1}$~cm$^{-2}$~s$^{-1}$, $\alpha=2.04$ and $E_{cut}=17.9$~TeV, predicts noticeably higher flux of neutrinos at the most relevant energies for the detection of neutrinos, $E_\gamma \geq 10$~TeV. This is visible in Fig.~\ref{fig:RXJ}, where the expected neutrino fluxes from both H.E.S.S. measurements are shown together with the flux sensitivities of the two instruments for a source with a radius of $0.6$~deg. High quality spectroscopic measurements of gamma rays at the highest energies are still missing, due to the limited sensitivity of current instruments. The uncertainty on the gamma-ray flux above $\sim10$~TeV is expected to be substantially diminished through CTA observations on the source, for a rather short time. Remarkably, even for the lowest predicted neutrino flux, a statistically significant detection of the latter by KM3NeT seems realistic for timescales of 10 years. This is well in agreement with the lack of events from RX J1713.7-3946 by the current generation of neutrino telescopes, whose upper limits on its neutrino flux amounts to $6.7\times 10^{-12}$~TeV cm$^{-1}$ s$^{-1}$ in the case of ANTARES \cite{Albert:2017ohr}, and to $9.2\times 10^{-12}$~TeV cm$^{-1}$ s$^{-1}$ for the IceCube detector \cite{Aartsen:2016oji}. Note that KM3NeT consists of a phased and distributed infrastructure, whose next-phase is the so-called ARCA configuration: an array made of two building blocks, whose completion is planned by 2020. In \ref{sec:arca} we investigate the possibility of a neutrino detection from the SNR with the ARCA detector, since it represents a timely study and permits a meaningful comparison with what foreseen by the KM3NeT collaboration, as reported in \cite{Adrian-Martinez:2016fdl}.

\subsection{The Galactic Center Ridge}
\label{sec:gcr}

Another promising object from the point of view of multi-TeV neutrino detection is the Galactic Center Ridge. The observations with the H.E.S.S. telescope revealed the presence of a diffuse emission component from a 200 pc region around the Galactic Center \cite{Aharonian:2006au, Abramowski:2016mir}. The hard energy spectrum of the emission extends well above $10$~TeV without any indication of a spectral break or cutoff \cite{Abramowski:2016mir}. If the emission is hadronic, neutrinos are expected as well and this source is another candidate to be detected by KM3NeT \cite{Celli:2016}. The H.E.S.S. measurements of the Galactic Center Ridge spectrum, as recently reported in \cite{hess2017}, point toward an unbroken power-law of the form:
\begin{equation}
\frac{dN_{\gamma}}{dE} (E) = \phi_0 \left( \frac{E}{E_0} \right)^{-\alpha}
\end{equation}
with $E_0=1$~TeV, $\phi_0=1.2\times 10^{-8}$~TeV$^{-1}$~cm$^{-2}$~sr$^{-1}$~s$^{-1}$, $\alpha=2.28$. This spectrum corresponds to the region $|l|\le1.0$~deg, $|b|\le0.3$~deg. The estimation of the sensitivities is done for the same sky region, namely a rectangular box with longitudinal size of 2.0~deg and latitudinal size of 0.6~deg. The results are shown in Fig.~\ref{fig:GCR}, where three neutrino spectra are calculated: the unattenuated power law that closely follows the gamma-ray measurements and two more spectra obtained assuming an exponential cutoff in gamma-rays at 100 TeV and 1 PeV.  One can see that only for the case for which the location of the gamma-ray cutoff energy is beyond 100 TeV, does the gamma-ray data guarantee a statistically significant detection of the counterpart neutrinos by KM3NeT.
The gamma-ray flux above 100 TeV is too weak to be detected by the H.E.S.S. telescopes, even after a decade of continuous monitoring of this region. The exploration of this energy domain requires a more powerful gamma-ray instrument such as CTA. This can be seen in Fig.~\ref{fig:GCR}. However, it is necessary to note that gamma-rays, even in the case of their effective production above 100 TeV,  hardly can escape the Ridge because of the absorption through interactions with the enhanced far infrared radiation fields in the central 100~pc region. Thus it is likely that the neutrinos remain the only messengers of information about the cosmic-ray protons with energies more than 1~PeV.  This opens a unique opportunity for KM3NeT to provide a major contribution to the exploration of the cosmic-ray PeVatron in the Galactic Center.

%%%%%%%%%%%%%%%%%%%%%%%%%%%%%%%%%%%%%%%%%%%''''
\section{The second HAWC catalog}
\label{sec:2hwc}
\begin{figure*}[!h]
\centering
\subfigure[]{\label{fig:hawcps} \includegraphics[height=0.28\textheight,width=0.49\textwidth]{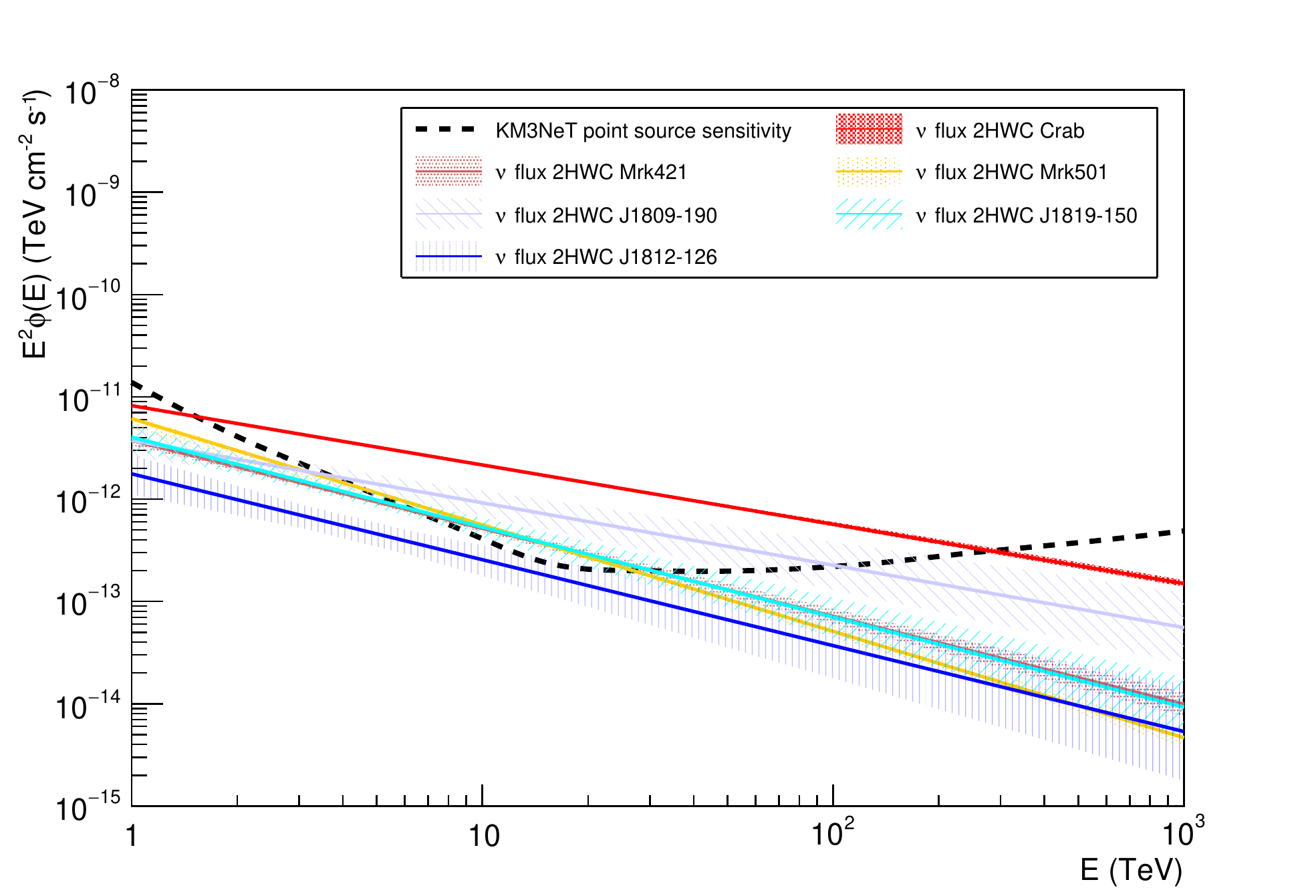}} 
\subfigure[]{\label{fig:hawc05} \includegraphics[height=0.28\textheight,width=0.49\textwidth]{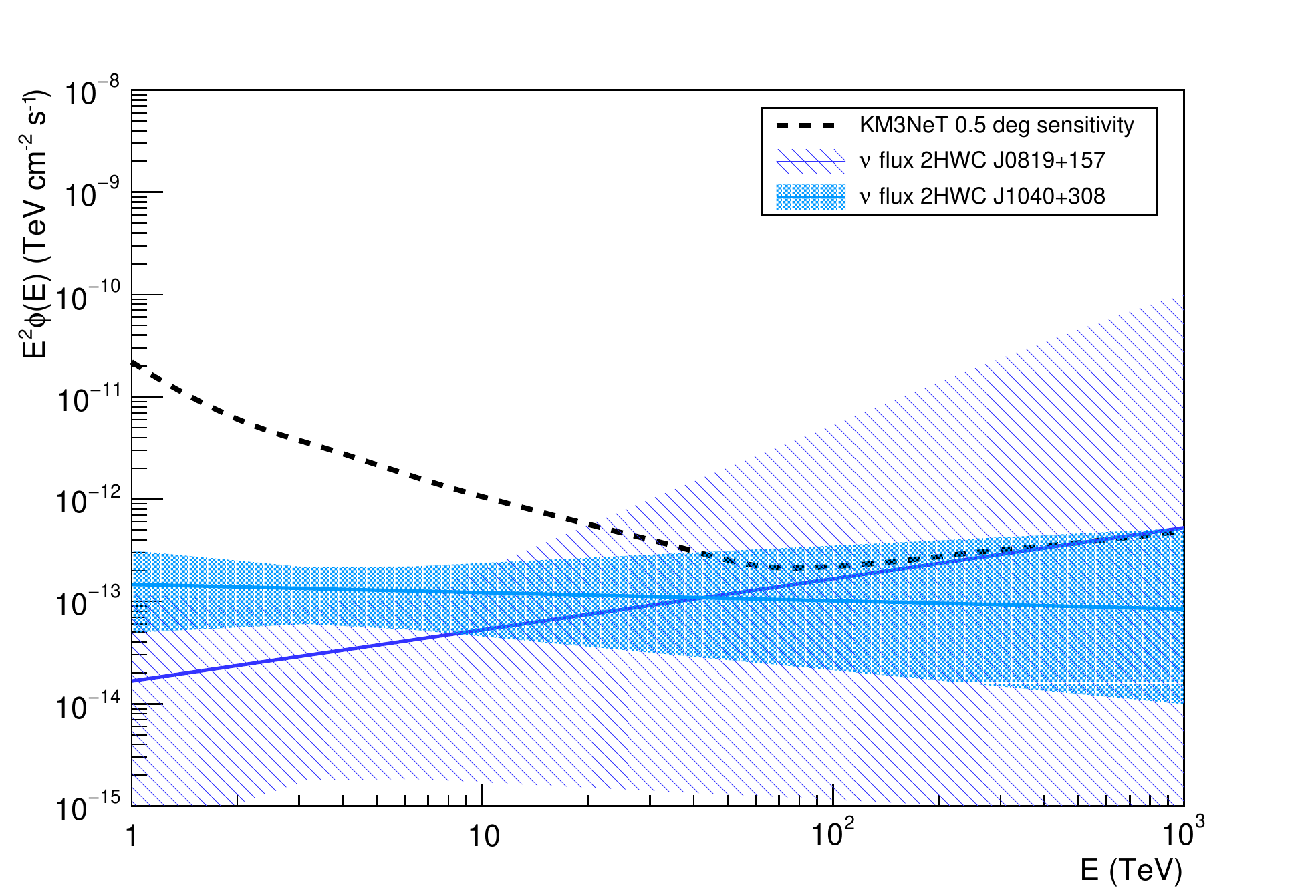}}
\subfigure[]{\label{fig:hawc06} \includegraphics[height=0.28\textheight,width=0.49\textwidth]{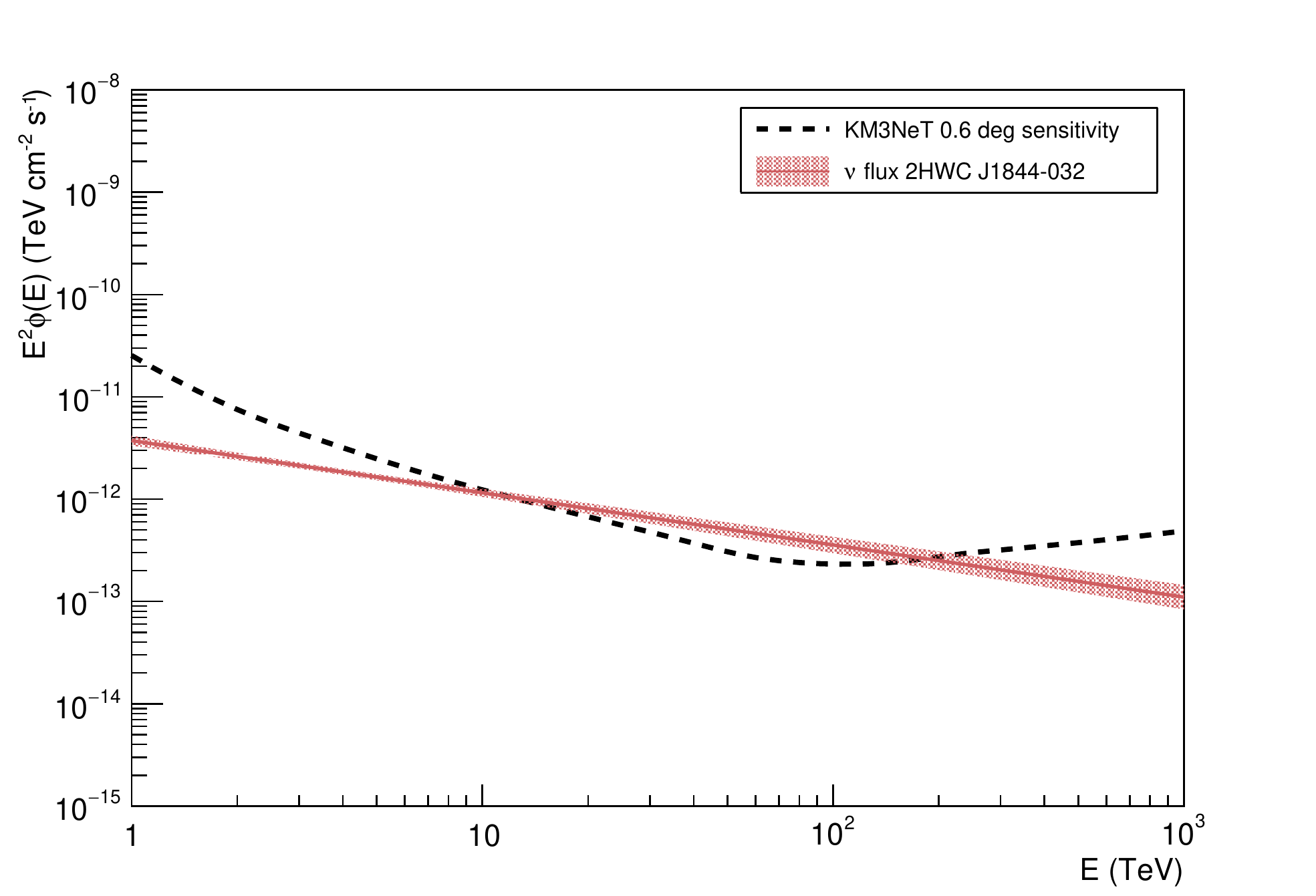}}
\subfigure[]{\label{fig:hawc07} \includegraphics[height=0.28\textheight,width=0.49\textwidth]{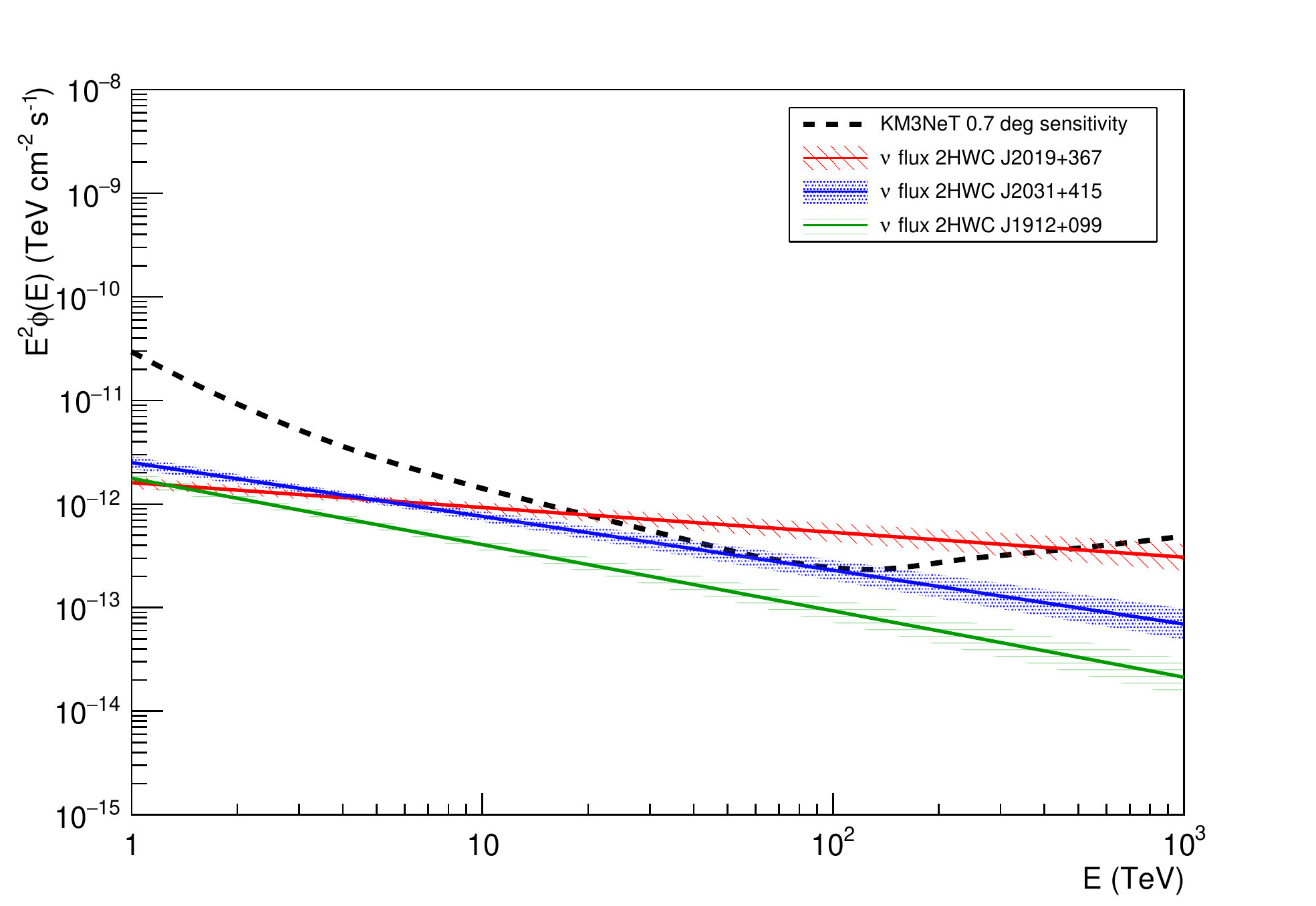}}
\caption{ \label{fig:hawc_1} Muon neutrino fluxes (solid lines) of the sources reported in the HAWC catalog \cite{Abeysekara:2017hyn} and KM3NeT minimum detectable flux (dashed line), as computed according to the procedure described in Sec.~\ref{sec:sens}, to: (a) point-like sources, (b) 0.5~deg extended sources, (c) 0.6~deg extended sources and (d) 0.7~deg extended sources.}
\end{figure*}

\begin{figure*}[!h]
\centering
\subfigure[]{\label{fig:hawc08} \includegraphics[height=0.28\textheight,width=0.49\textwidth]{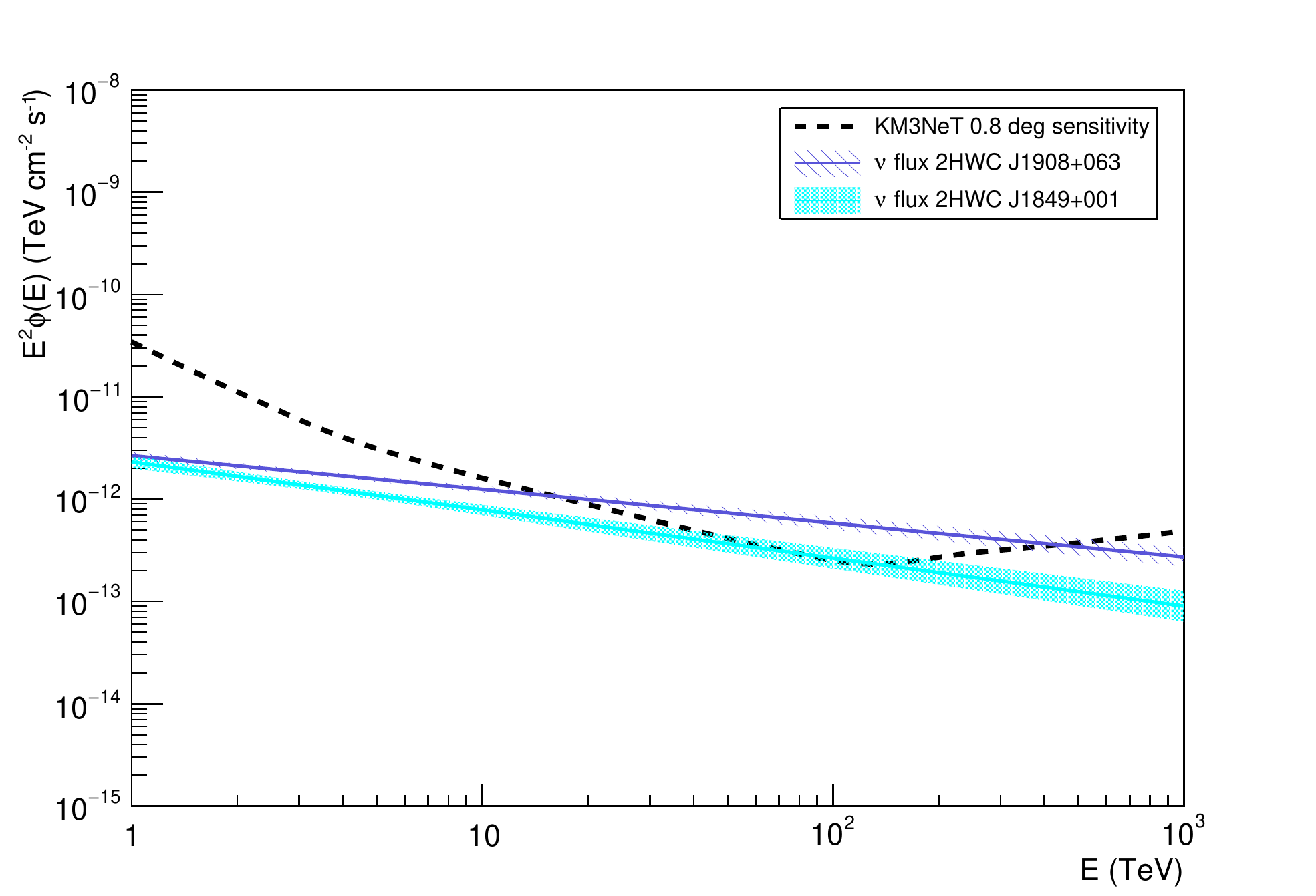}}
\subfigure[]{\label{fig:hawc09} \includegraphics[height=0.28\textheight,width=0.49\textwidth]{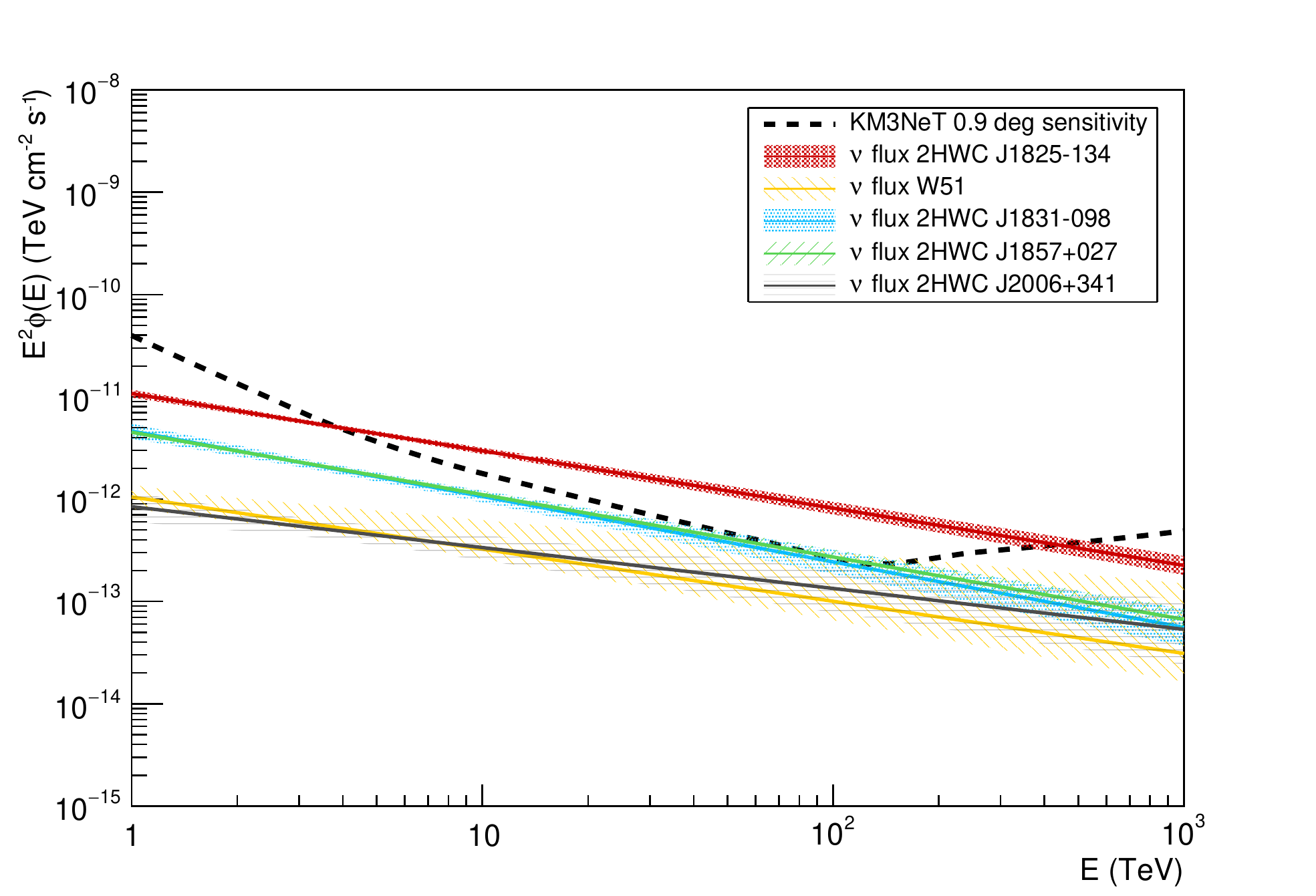}}
\subfigure[]{\label{fig:hawc1} \includegraphics[height=0.28\textheight,width=0.49\textwidth]{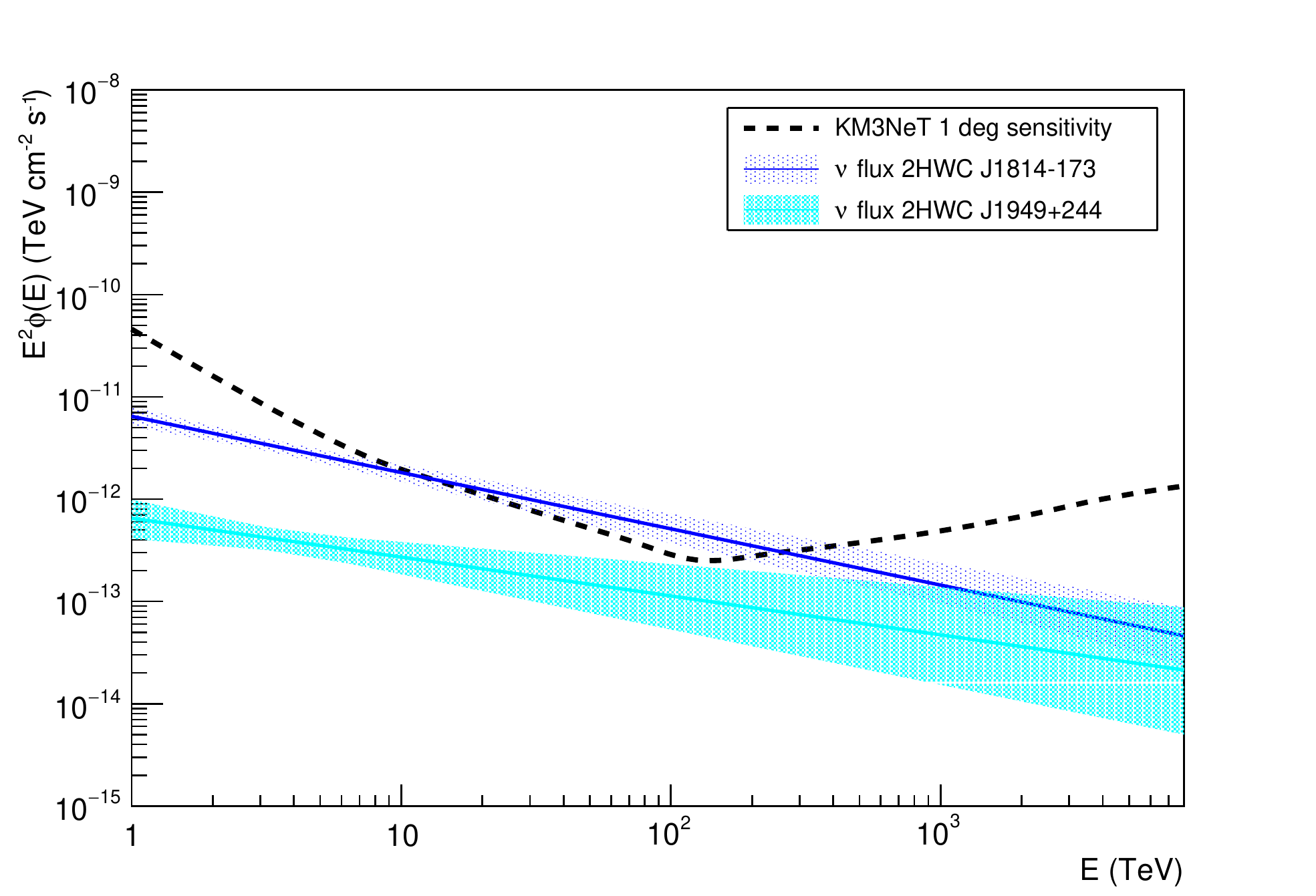}}
\subfigure[]{\label{fig:hawc2} \includegraphics[height=0.28\textheight,width=0.49\textwidth]{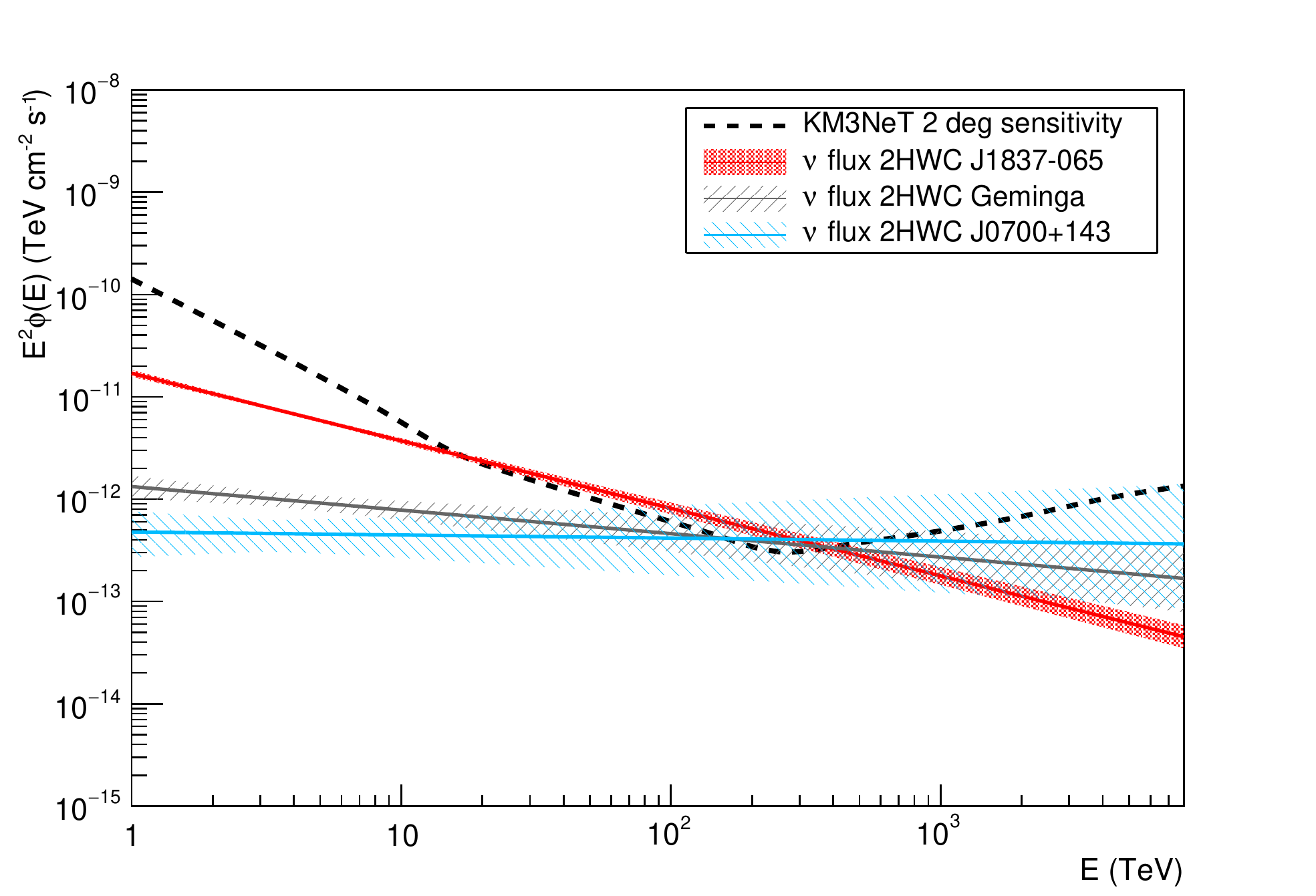}}
\caption{ \label{fig:hawc_2} Muon neutrino fluxes (solid lines) of the sources reported in the HAWC catalog \cite{Abeysekara:2017hyn} and KM3NeT minimum detectable flux (dashed line), as computed according to the procedure described in Sec.~\ref{sec:sens}, to: (a) 0.8~deg extended sources, (b) 0.9~deg extended sources, (c) 1.0~deg extended sources and (d) 2.0~deg extended sources.}
\end{figure*}

%%%%%%%%%%%%%%%%%%%%%%%%%%%%%%%%%%%%%%%%%%%''''
Beyond the search for neutrino emission from individual sources, population studies are expected to provide a deep comprehension of the acceleration mechanism acting at the source. We analyze here the recently published second High Altitude Water Cherenkov (HAWC) catalog of TeV sources, as reported in \cite{Abeysekara:2017hyn}, discussing the capability for neutrino telescopes to investigate such fluxes. A total of 39 VHE gamma-ray sources are reported in the catalog, both galactic and extra-galactic: of these, two are associated with blazars, two with SNRs, seven with pulsar wind nebulae (PWNe). Of the rest, 14 have possible associations with SNR, PWN, and molecular clouds, while the remaining 14 are still unidentified. With respect to other TeV catalogs from IACT arrays, a HAWC extended source is actually a sky region where more sources might be overlapping, given the poor angular resolution with respect to IACT. We consider all the emission coming from such a region as a single source, given that neutrino telescopes have a quite similar angular resolution to water-based Cherenkov gamma-ray instruments. Several sources reported in \cite{Abeysekara:2017hyn} are tested under different angular extension hypotheses, leading to different spectral fits: the spectral fit corresponding to the more extended source assumption is considered here. Furthermore, Geminga is flagged twice in the HAWC catalog (both as a point-like and as a 2~deg extended object): therefore, the final list counts a total of 40 gamma-ray emitters.
In the view of a neutrino detection, the sensitivity of the KM3NeT detector to upward-going track events for different angular extension of HAWC sources has been studied. Although this is not the best source sample to be investigated with KM3NeT through upgoing muons, given their sky position, nonetheless we here consider it since it is interesting to exploit sources not previously detected in the TeV energy region. The sensitivity of KM3NeT is reported in Fig.~\ref{fig:hawc_1} and Fig.~\ref{fig:hawc_2}, together with the expected neutrino fluxes. The case of point-like sources with fluxes in the reach of KM3NeT is reported as well. In particular Fig.~\ref{fig:hawc_1} shows the neutrino expectations for: \ref{fig:hawcps} HAWC point-like sources; \ref{fig:hawc05} for an extension of 0.5~deg; \ref{fig:hawc06} for 0.6~deg; Fig.~\ref{fig:hawc07} for 0.7~deg. Analogously \ref{fig:hawc08} shows the results for 0.8~deg sources; \ref{fig:hawc09} for 0.9~deg sources; \ref{fig:hawc1} for 1.0~deg sources and \ref{fig:hawc2} for 2.0~deg sources. We here recall that the computation of neutrino spectra are realized without accounting for possible absorption of gamma-rays, which might be relevant for extra-galactic sources: in this case, neutrino fluxes would increase since neutrinos do not suffer of absorption. As visible in Fig.~\ref{fig:hawc_1} to \ref{fig:hawc_2}, promising sources for a neutrino detection are represented by 2HWC J1809-190, 2HWC J1819-150, Crab, Mrk421, 2HWC J1844-032, 2HWC J2019+367, 2HWC J1908+063, 2HWC J1825-134, 2HWC J1814-173 and 2HWC J1837-065.
Since we are considering only sources below the horizon, the source visibility needs to be taken into account. The sources with more than 50\% visibility at the KM3NeT latitude are constituted by 2HWC J1809-190, 2HWC J1819-150, 2HWC J1814-173, 2HWC J1825-134, 2HWC J1844-032 and 2HWC J1837-065. The inclusion of down-going events, with specific analysis features allowing the reduction of the atmospheric muon background, will permit to include in the analysis the whole data-taking period: in this case, however, the selection efficiency would further reduce the triggered sample by a factor of a few. Finally, in the case of two degree extended sources, the cascade channel might be added, given that the source dimensions are comparable to the shower angular resolution at high energies (about $2^{\circ}$ on average).
A combined track and cascade analysis is thus the most effective strategy to pursue the goal of identifying neutrino sources in case of very extended objects.

\section{Discussion and conclusions}
\label{sec:disc}
In this work we investigated the discovery potential of extended sources by the future KM3NeT, in relation to the constraining power of the next-generation gamma-ray telescope, CTA. Unless multi-TeV photons are absorbed inside the sources or during their propagation through the interstellar or intergalactic radiation fields, gamma-ray observations can safely be considered a powerful tool to explore the potential for finding astronomical neutrino sources. In order to accomplish this purpose, we here explored the sensitivities of both instruments to extended sources.  
We here arrived to the following conclusions:
(i) the sensitivity to extended sources shows a degradation with increasing source angular size such that it is maximum at low energies, reducing at intermediate energies and tending to disappear at very high energies; (ii) the most important energy region for the detection of neutrino sources is above $10$~TeV. In this energy region, we see no strong dependence of the CTA minimum detectable gamma-ray flux with source size, with a comparison of the performances of the two instruments showing that above this energy a joint exploration of the VHE sky in gamma rays and neutrinos will be possible. Nowadays, we do have gamma-ray observations above $10$~TeV, as the surveys of the Galactic Plane realized by the current ground-based telescopes, HAWC and H.E.S.S. These observations are already constraining from the point of view of their counterpart neutrino detection. However, still room is available for the presence of Galactic neutrino emitters. In the near future CTA will explore the entire Galactic Plane in the 10-100 TeV energy domain, which is critical for Galactic sources, since most of them appear to have a spectrum with a cut-off around an energy of 10-50 TeV. Therefore, CTA will significantly reduce the limits on the neutrino source expectations, setting conclusions on possible Galactic neutrino emitters through its extremely strong constraining power.  \\
Our analysis shows that, assuming a source emitting a gamma-ray $E^{-2}$ differential energy spectrum through a fully hadronic mechanism, a minimum gamma-ray flux of $E^2 \phi_\gamma(10~\textrm{TeV})>1 \times 10^{-12}$~TeV~cm$^{-2}$~s$^{-1}$ is necessary in order for its neutrino counterpart to be detectable with a 3$\sigma$ significance on a time scale of 10 years with KM3NeT. This result assumes that the source has an angular size of $R_{src}=0.1$~deg. In the extreme case of a source with a radial extent of $R_{src}=2.0$~deg, only sources brighter than $E^2 \phi_\gamma(10~\textrm{TeV})>2 \times 10^{-11}$~TeV~cm$^{-2}$~s$^{-1}$ will be within the reach of neutrino telescopes. These estimates (obtained according to Eq. 34 in \cite{Villante:2008qg}) are very weakly dependent on the source spectral index and are consistent with previous evaluations performed in \cite{Vissani:2011vg} for the case of point-like sources. \\
In particular, RX J1713.7-3946 and the Galactic Center Ridge remain potential sources for neutrinos. We found that a decade of observations is required for a $3\sigma$ (in each energy bin) neutrino detection from the SNR and from the most optimistic set of parameters considered for the Galactic Center Ridge (a cut-off in the gamma-ray spectrum at energies above 100~TeV).

\bigskip
\noindent
We would like to emphasize that the instrument performances here presented come as an original study done by the authors: more sophisticated analysis approaches tailored to the detection of extended objects might be developed by the  CTA and KM3NeT Collaborations, leading to improved sensitivities with respect to that found in this work.
However, in the energy domain relevant for a multi-messenger observation of VHE extended sources (i.e. above 10~TeV), the differences with the results here reported are not expected to be significant. This is due to the signal detection being determined by the limited statistics, with the main uncertainty residing in the knowledge of the instrument collection area.

\appendix
\setcounter{figure}{0}
\renewcommand{\thetable}{A\arabic{figure}}
%%%%%%%%%%%%%%%%%%%%%%%%%%%%%%%%%%%%%%%%%%%''''
\section{Off-axis sensitivity}
\label{sec:offaxis}
%%%%%%%%%%%%%%%%%%%%%%%%%%%%%%%%%%%%%%%%%%%''''

\begin{figure}[ht]
\centering
\includegraphics[width=0.49\textwidth]{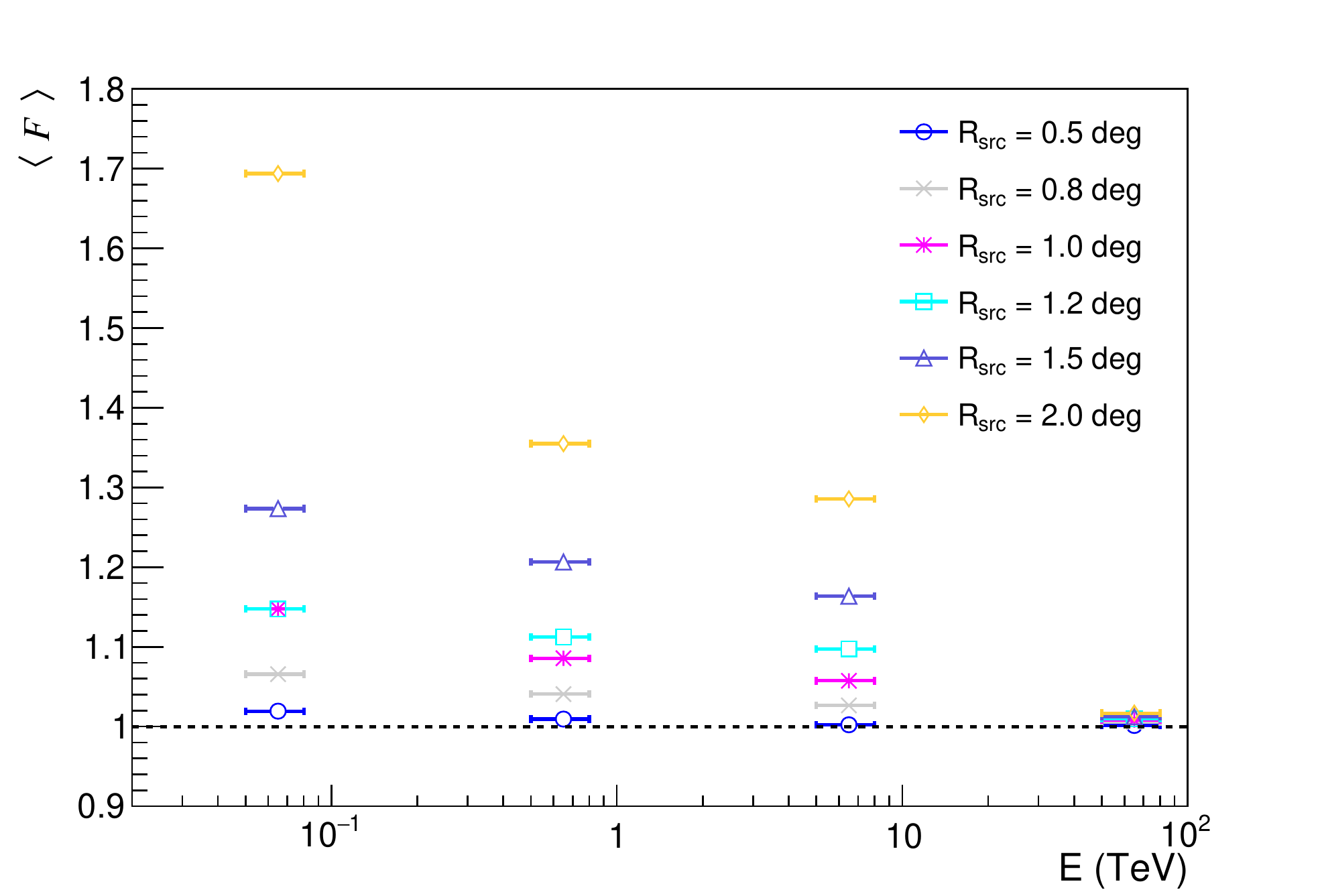}
\caption{CTA South correction factor $\langle \mathcal{F} \rangle$ to account for the off-axis degradation of the instrument sensitivity to extended sources. See the text for a detailed description of the factor $\langle \mathcal{F} \rangle$, as defined in Eq.~\ref{eq:cf}. The widths of the energy bins correspond to the intervals for which the CTA point-source off-axis estimations are available.}
\label{fig:OffAxisSens}
\end{figure}

The sensitivity study performed through this paper assumes that the source is located in the center of the CTA FoV. However, in the case of extended sources, a degradation of the telescope performances should be accounted for, since part of the source would result displaced from the FoV center. \\
In this Section we estimate a correction factor to be applied to the CTA sensitivity curves shown in Sec.~\ref{sec:sens}, in order to take into account the degradation of the instrument response due to the offset of the source. We consider the publicly available results of the CTA Southern array for what concerns the point-source detectability off-axis, as reported in \cite{CTAperformaceWeb}. Here, the instrument point-source off-axis sensitivity relative to the one at the center of the FoV is presented in four different energy bins (i.e. $50-80$~GeV, $0.5-0.8$~TeV, $5-8$~TeV and $50-80$~TeV). In the same energy intervals we evaluate the correction factor $\langle \mathcal{F} \rangle$ as the average value of the CTA expectations over the total extension of the objects. The objects with total radius $R_{ROI}$ (see Eq.~\ref{eq:rroi}) are treated as composed by concentric annuluses with size $2\sigma_{PSF}$, where $\sigma_{PSF}$ is the mean value of the angular resolution in each of the four energy bins. Opting for a conservative approach, we consider a set of $N$ annuluses, with $N$ the smallest integer such that the radius of the source can be expressed as a finite multiple of the instrument angular resolution, so that $N \cdot (2\sigma_{PSF}) \ge R_{ROI}$. The contribution of each annulus to $\langle \mathcal{F} \rangle$ is then weighted according to the area of the ring $A_{ring}$ itself, yielding to the definition of $\langle \mathcal{F} \rangle$ as:
\begin{equation}
\label{eq:cf}
\langle \mathcal{F} \rangle = \frac{\sum_{i=1}^{N} f_{i} \cdot A_{ring,i} }{ \sum_{i=1}^{N} A_{ring,i} }
\end{equation}
where $f_{i}$ is the value of relative worsening at a fixed distance from the camera center, as inferred from \cite{CTAperformaceWeb}. In Fig.~\ref{fig:OffAxisSens} the estimated correction factor $\langle \mathcal{F} \rangle$ is shown for sources with an extension $\ge0.5$\,deg, as for smaller distances to the camera center the telescope sensitivity does not suffer any significant worsening (i.e. $\langle \mathcal{F} \rangle =1$). The worsening of sensitivity is less than a factor of two, even for the largest source extension considered in this paper, i.e. $2$~deg, in the lowest energy interval.

%%%%%%%%%%%%%%%%%%%%%%%%%%%%%%%%%%%%%%%%%%%''''
\section{The 4 years ARCA sensitivity}
\label{sec:arca}
%%%%%%%%%%%%%%%%%%%%%%%%%%%%%%%%%%%%%%%%%%%''''

\begin{figure}[ht]
\centering
\includegraphics[width=0.49\textwidth]{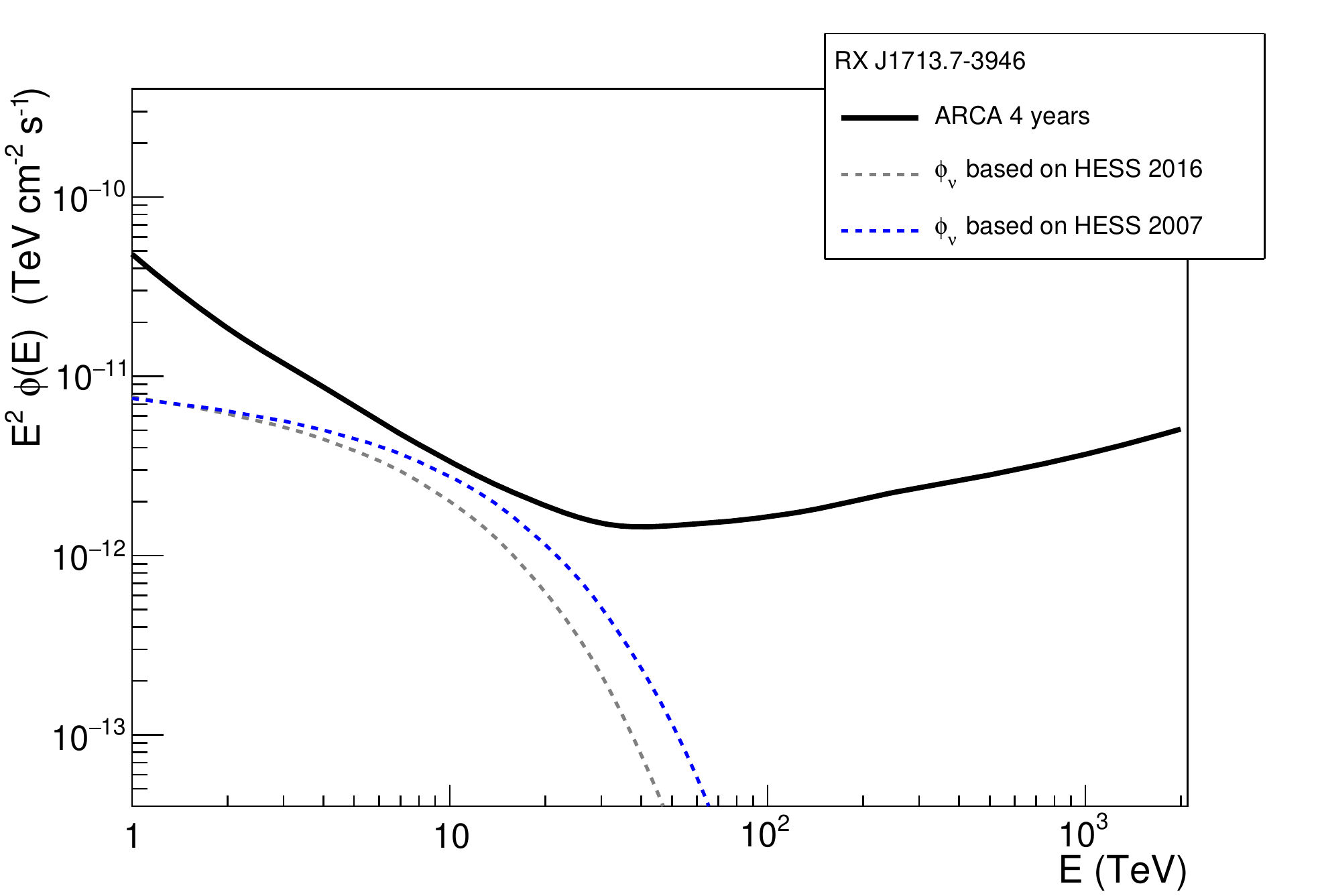}
\caption{The 4 years ARCA minimum detectable flux (estimated according to the procedure described in Sec. \ref{sec:sens}) for a source of angular extension equal to $R_\textrm{src}=0.6$~deg, shown as solid black line. The dashed curves are the binned muon neutrino fluxes computed according to the model in \cite{Villante:2008qg}.}
\label{fig:RXJarca}
\end{figure}

In this final Section, we explore the possibility of a neutrino detection from the extended source RX J1713.7- 3946 (described in Sec.~\ref{sec:sources}) on a short-term time scale through the ARCA detector: this infrastructure will be composed by two KM3NeT building blocks, for an effective volume of a cubic kilometer, representing the next-phase (the so called Phase-2.0) of the KM3NeT project \cite{Adrian-Martinez:2016fdl}. As stated in \cite{Adrian-Martinez:2016fdl}, the neutrino emissions from SNR RX J1713.7- 3946 might be investigated at $3\sigma$ significance in 4 years of observations through ARCA. The estimation of the differential sensitivity of such an instrument for the detection of a neutrino signal from the extended region of the SNR RX J1713.7-3946 has been performed according to the procedure described in Sec.~\ref{sec:sens}, therefore considering the source angular extension equal to $R_\textrm{src}=0.6$~deg. In order to show a comparison with the expectations reported in \cite{Adrian-Martinez:2016fdl}, an observation time of 4 years is here assumed. The resulting minimum detectable flux is shown in Fig.~\ref{fig:RXJarca}, together with the binned muon neutrino fluxes from the source calculated according to the model in \cite{Villante:2008qg}.
The result in \cite{Adrian-Martinez:2016fdl} is well consistent with our evaluation: when considering integrated sensitivity studies, a $3\sigma$ neutrino detection from the SNR is an achievable goal with 4 years of data taking. However, a spectrometric study of the neutrino emission form this source would require a larger instrumented volume as well as a longer exposure.

\section*{Acknowledgement}
\vspace{-0.2cm}
We thank G.~Morlino and F.~Vissani for useful discussions. We are also grateful to R.~Coniglione, L.~A.~Fusco, P.~Sapienza, A.~Trovato, R.~Zanin, E.~de~O\~na~Wilhelmi and A.~Taylor for fruitful comments.

\vspace{0.2cm}
\noindent
This paper has gone through internal review by the CTA Consortium.

\vspace{1.0cm}
\bibliographystyle{elsarticle-num} 
\bibliography{./biblio}

\end{document}